\documentclass[11pt,a4paper]{article}

\usepackage[round, sort]{natbib}

\usepackage{ifpdf}
\usepackage{color}
\usepackage{latexsym}                   
\usepackage{epsfig}                     
\usepackage{graphicx}
\usepackage{amssymb, amsmath, amsthm,bbm}   
\usepackage{enumerate}                  
\usepackage{mathrsfs}                   
\usepackage{geometry}                                                                          
\usepackage{verbatim}                                                                           
\usepackage{tabularx}     

\usepackage{setspace}

\newtheorem{thm}{\bf Theorem}[section]
\newtheorem{rem}{\bf Remark}[section]
\newtheorem{prop}{\bf Proposition}[section]

\newcommand{\PP}{\mathbb P}
\newcommand{\EE}{\mathbb E}
\newcommand{\NN}{\mathbb N}

\begin{document}

\title{Ages, sizes and (trees within) trees of taxa and of urns,
from Yule to today}

\author{%
Amaury Lambert\\
Stochastic Models for the Inference of Life Evolution (SMILE)\\
Institute of Biology of ENS (IBENS)\\
 CNRS, INSERM, Universit\'e PSL\\
 \'Ecole Normale Sup\'erieure\\
 46 rue d'Ulm 75005 Paris\\
 \&\\
Center for Interdisciplinary Research in Biology (CIRB)\\
CNRS, INSERM, Universit\'e PSL\\
Coll\`ege de France\\
11 place Marcelin Berthelot 75005 Paris\\ 
}

\maketitle

\begin{abstract}

The paper written in 1925 by G. Udny Yule that we celebrate in this special issue introduces several novelties and results that we recall in detail. 

First, we discuss Yule (1925)'s main legacies over the past century, focusing on empirical frequency distributions with heavy tails and random tree models for phylogenies. We estimate the year when Yule's work was re-discovered by scientists interested in stochastic processes of population growth (1948) and the year from which it began to be cited (1951, Yule's death). We highlight overlooked aspects of Yule's work (e.g., the Yule process of Yule processes) and correct some common misattributions (e.g., the Yule tree).

Second, we generalize Yule's results on the average frequency of genera of a given age and size (number of species). We show that his formula also applies to the age $A$ and size $S$ of any randomly chosen genus and that the pairs $(A_i, S_i)$ are equally distributed and independent across genera. This property extends to triples $(H_i, A_i, S_i)$, where $(H_i)$ are the coalescence times of the genus phylogeny, even when species diversification within genera follows any integer-valued process, including species extinctions. Studying $(A, S)$ in this broader context allows us to identify cases where $S$ has a power-law tail distribution, with new applications to urn schemes.
\end{abstract}

\section{Introduction}

We first provide the reader with some basic notions required to read this introduction.

\subsection{Some preliminary material}

A \textbf{\emph{birth--death process}} is a Markov chain in continuous time which lives in the natural integers and only jumps by $\pm 1$. 

The process $(Z_t;t\ge 0)$ counting the size of a population where particles divide at rate $\lambda$ and die at rate $\mu$ independently is a birth-death process called the \emph{linear birth--death process}, and sometimes even \emph{the} birth--death process, as we will do here. 

At any time $t$ when $Z_t=n$, the probability that $Z_t$ jumps by $+1$ in the infinitesimal interval $(t,t+dt)$ equals $n\lambda\, dt$, and the probability that $Z_t$ jumps by $-1$ equals $n\mu\, dt$. Equivalently, $Z_t$ jumps after waiting an exponential time with parameter $n(\lambda+\mu)$, at the end of which the probability of a positive jump (jump by $+1$) is $n\lambda/(n\lambda+n\mu)=\lambda/(\lambda+\mu)$, and the probability of a negative jump (jump by $-1$) is $\mu/(\lambda+\mu)$, independently of the exponential waiting time.

When $\mu=0$ the birth-death process is called \textbf{\emph{Yule process}} (sometimes Yule-Furry process), or \textbf{\emph{pure-birth process}}, and indeed was introduced by G. Udny Yule.

\begin{figure}[h] 
\begin{center} 
\includegraphics[width=.8\textwidth]{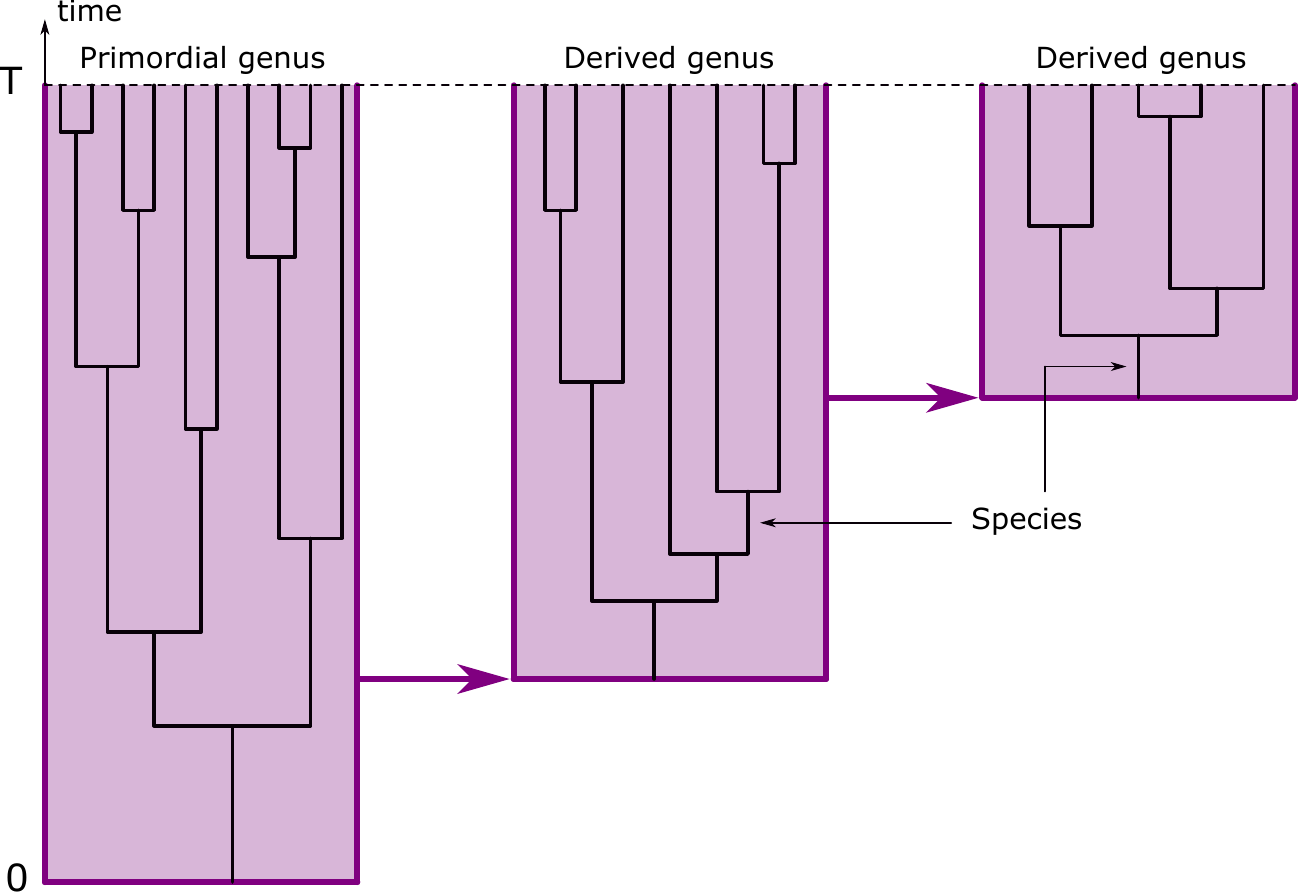} 
\end{center}
 \caption{\textbf{The \cite{Y} model is a pure-birth process (genera) of pure-birth processes (species).} Each genus gives birth at rate $g$ and upon birth, the daughter genus starts with one single species. Within each genus independently, species independently give birth at rate $s$. Here the process starts with one genus at time 0 and is stopped at time $T$.} 
 \label{fig:yiy}
 \end{figure}

In his seminal paper, \cite{Y} actually considers a much more complex model: pure-birth processes modeling species diversification nested in a pure-birth process modeling genus diversification (see Figure \ref{fig:yiy}), namely: \textbf{each genus gives birth at rate $g$ independently to a new genus initially containing one species, and in each genus independently, species divide according to a pure-birth process with rate $s$}.

As we will see, this process leads to the study of a distribution $(f_n^Y;n\in \NN)$ on the positive integers, known as the \textbf{\emph{Yule distribution}} (sometimes Yule-Simon distribution):
\begin{equation}
\label{eqn:integral-form}
f_n^Y=\int_0^\infty ge^{-gt}e^{-st}\left(1-e^{-st}\right)^{n-1}dt\qquad n\ge 1.
\end{equation}
It is well-known that
\begin{equation}
\label{eqn:yule-simon}
f_n^Y= \frac{g}{g+s n}\prod_{k=1}^{n-1} \frac{s k}{g+s k} = \frac{\alpha}{\alpha+ n}\prod_{k=1}^{n-1} \frac{ k}{\alpha+ k} = \frac{\alpha(n-1)!}{\prod_{k=1}^{n}(\alpha+k) }
=\frac{\alpha\Gamma (n)\Gamma(1+\alpha)}{\Gamma(n+1+\alpha)},
\end{equation}
where 
$$
\alpha:=g/s
$$
and $\Gamma$ is the standard Gamma function, so that as $n\to\infty$
\begin{equation}
\label{eqn:power-law}
f_n^Y\sim \alpha \Gamma (1+\alpha)\, n^{-\alpha-1}.
\end{equation}

\subsection{The main two legacies of \cite{Y}}

George Udny Yule (1871--1951) was a famous statistician, fellow of the Royal society (1922) and author of a renowned textbook ``An introduction to the theory of statistics'' \citep{yule1919introduction}, first published in 1911 and reprinted for the last and fourteenth time in 1950, see e.g., Chapter 15 of \cite{Bacaer}.\\

The 60-page article we are celebrating in the present special issue has had two distinct, important legacies in the elapsed century. In 2024, a query of publications citing \cite{Y} indeed shows that the $> $2500 available references to this article fall very clearly into one of the two distinct categories of topics:
\begin{itemize}
\item \textbf{Empirical frequency distributions with heavy tail}, in statistical physics, network theory, systems biology, economy and other social sciences;
\item \textbf{Random tree models for phylogenies}, in population genetics, phylogenetics, paleontology and mathematics applied to these fields.  
\end{itemize} 
Citations of the first kind were initiated in 1955 by a rediscovery of Yule's work by Herbert A. Simon
 (who would receive in 1978 the Nobel Memorial Prize in Economic Sciences) and his interpretation in terms of `preferential attachment', as will be coined later by \cite{barabasi1999emergence} for their study of a random graph with vertices of unusually large degree. This line of research has led to an abundance of papers fitting count distributions to Pareto, Zipf or other heavy-tailed distributions illustrating the so-called `rich get richer' effect (sometimes also called the `Matthew effect'). These works cite \cite{Y}, \cite{simon1955class}, and later \cite{barabasi1999emergence}, as the seminal works on this topic, to the point that some authors erroneously attribute the first random graph model to Yule...

In the models of \cite{Y} and \cite{simon1955class}, which actually differ (see Section \ref{sec:urns}), a growing ensemble of collectives (genus, urn) containing particles (species, balls) is considered. At each step of this growth process, either a new urn starting with one ball is added or one urn, picked in proportion to the number of balls it contains, receives an extra ball. After $n$ steps of this process, there are $n$ balls randomly partitioned into a random number of urns and the question of interest is to compute and study the frequency $f_k$ of urns with $k$ balls, in particular when $n$ is large. This `preferential attachment' scheme yields frequency distributions where $f_k$ has a power-law tail, embodying the existence of urns of abnormally large size (i.e., number of balls). \cite{simkin2011re} review a quantity of papers studying processes leading to power-law distributions involving species, urns, graphs starting with works by (Yule and) Willis. They provide short proofs of the different results known at the time and relate them when possible.

Later in his career, Yule has been interested in the empirical frequencies of words in literary works, one of the statistical patterns that also triggered the research of H. A. Simon on `skew' distributions. Yule even wrote an entire book about this topic \citep{yule1944statistical}, but as Simon noticed \citep[p. 439]{simon1955class}:
\begin{quote}
``It is interesting and a little surprising that when Yule, some twenty years after this discovery, examined the statistics of vocabulary, he did not employ this model to account for the observed distributions of word frequencies. Indeed, in his fascinating book on ``The Statistical Study of Literary Vocabulary'' (1944) he nowhere refers to his earlier paper on biological distributions''.
\end{quote}

Citations of the second kind started in the early 50's. Between 1925 and 1950, the works of botanist James Small account for the near totality of available references to \cite{Y} and are mainly concerned with testing quantitative evolutionary theories against data, starting with \cite{small1939quantitative}. The rest of the scientific community seems to have ignored \cite{Y} until 1948, as we will now see.

The pure-birth process has been reinvented in a particle physics paper\footnote{Available references to \cite{furry1937fluctuation} are in their vast majority concerned with multielectron cascades and cosmic ray showers. Very few works in population biology or probability theory seem to be aware of this paper (but see \citealt{bartlett1951use, kendall1949stochastic, ramakrishnan1951some, goodman1967probabilities, whittle1952certain}), to the notable exception of a number of textbooks dealing with stochastic processes or mathematical ecology \citep{harris1963theory, kot2001elements, parzen1999stochastic, ricciardi1986stochastic, beichelt2018applied, ramakrishnan1959probability, hopcraft2014dynamics}, which by nature devote more space to the historical side of science.} by \cite{furry1937fluctuation}, hence the name sometimes heard of `Yule-Furry process'. Then, a paper written in German by the great probabilist William Feller \citep{feller1939grundlagen}, which title could be translated as ``The basics of the Volterran theory of the struggle for existence in a probabilistic treatment'' introduces the birth-death process, and David G. Kendall \citep{kendall1948generalized}, today also famous for his theory of shape \citep{kendall1977diffusion,kendall1989survey} and future PhD advisor of J.F.C. Kingman, generalizes it to time-inhomogeneous rates.
In parallel, the Danish physicist Niels Arley had defended his doctoral thesis entitled ``On the theory of stochastic processes and their application to the theory of cosmic radiation'', published later as \cite{arley1948theory} and obviously aware of \cite{furry1937fluctuation}, where he had independently introduced and studied the birth-death process, also mentioned as Example (10.IV) by \cite{arley1944theory}. \cite{arley1949birth} also considers time-inhomogeneous birth-death processes in a short paper where he simplifies an argument of \cite{kendall1948generalized}.

Although G. Udny Yule had introduced the pure-birth process in 1925, none of these papers published between 1937 and 1949 cite his work, which seems to have been rediscovered only at the occasion of the 1948 ``Symposium on Stochastic Processes'' organized by the Royal Statistical Society. A tremendously instructive account of this symposium gathering the community of mathematicians and physicists studying the growth of populations (in particular Kendall, Bartlett, Moyal, Arley), is given by \cite{kendall1949stochastic}, who writes:
\begin{quote}
``Dr. Irwin, speaking from the Chair, has recalled some of the pioneer workers in this field, whose results have often been re-discovered and described in the new terminology. I have been very much struck, at all stages of the investigation, by the number of such surprises for a while concealed in the literature. I am sure there must be many more references to early work which I have omitted in ignorance; I regret none more than that to a paper of Mr. Yule containing a derivation of what has since been called the Furry process.'' 
\end{quote}
If one clearly sees the filiation between \cite{furry1937fluctuation} and \cite{arley1948theory}, it remains unclear after reading \cite{kendall1949stochastic} whether Irwin, and else who, brought \cite{Y} to the knowledge of Kendall and his colleagues during the symposium.

A couple of years\footnote{It seems that 15 years later, David G. Kendall would lose memory of his rediscovery: contrary to \cite{bartlett1951use} who did, \cite{kendall1966branching} does not cite \cite{Y} in his history of branching processes (see also \citealp{PMacP} in the same issue).} after the symposium appeared the first two papers citing \cite{Y} for his introduction of pure-birth processes \citep{moran1951estimation, bartlett1951use}, 26 years after this seminal paper and sadly enough, precisely the same year Yule died. It seems that Yule learnt about this late celebrity a few weeks before his death from the mouth of his friend and colleague at St John's College at Cambridge, Maurice G. Kendall (not to be confused with David G. Kendall), who testified in his obituary of Yule \citep{kendall1952george}:
\begin{quote}
``It was Yule who gave, in the "Introduction"\footnote{Yule's best-seller ``An introduction to the theory of statistics'' \citep{yule1919introduction}.}, formulae on correlated sums which are still being rediscovered by students of systematic sampling. It was Yule who invented the correlogram, though he did not invent the name; and likewise it was he who developed the autoregressive series, though again, another invented the name. It was Yule who cut through several pages of Pearsonian algebra to point out that the sampling formulae for partials must be of the same form as those of total correlation coefficients in normal variation, and hence paved the way for Fisher's derivation of the distribution of partial correlations. Only in one respect has his name been attached to a statistical concept, the so-called Yule process; and if I had not chanced to mention it in casual conversation a few weeks before his death he would have died in ignorance of the fact.''
\end{quote}


In Section \ref{sec:modernity}, we recall Yule's original ideas and results to demonstrate the modernity of his work by quoting some selected passages of his 1925 paper. In passing, we provide a glimpse at a few gaps that we will later intend to fill in Section \ref{sec:rejuvenation} by answering four questions displayed at the very end of Section \ref{sec:modernity}, p.\pageref{rem:q}.

\section{Modernity of \cite{Y}}
\label{sec:modernity}

Let us first illustrate our claim with a few excerpts of \cite{Y} showing its pioneer character.

\subsection{Some revolutionary ideas and findings}

The central object of Yule's paper is the pure-birth process, a particular form of what is known in modern terms as a \textbf{time-continuous Markov chain}. It is striking that Yule defines the pure-birth process using the notion of rate without naming it, 15 years before \cite{feller1940integro} proposed the first rigorous construction of time-continuous Markov chains: 
\begin{quote} 
[p. 33] 
``Let the chance of a species `throwing' a specific mutation, i.e., a new species of the same genus, in some small assigned interval of time be and suppose the interval so small that $p^2$ may be ignored compared with $p$.\\
We must now proceed to the limit, taking the time-interval $\Delta t$ as indefinitely small but the number of such intervals $n$ as large, so that the time $n\cdot \Delta t = t$ is finite. We
may write
$$
p = s \cdot \Delta t\qquad \qquad pn = st
$$
and we have the usual approximation
$$
q^n = (1-p)^n = (1-st/n)^n \sim e^{-st}.
$$
(...) that is, if $f_1$ is the proportional frequency of monotypic genera at time $t$, 
$$
f_1 = e^{-st}.''
$$
\end{quote} 
At the end of his paper, Yule even extends his method to include species deaths, in a passage that foreshadows the invention of the \textbf{birth-death process}:     
\begin{quote} 
[p. 78] ``...where $p$ is the proportion of survivors and $r$ gives the free rate of increase [on a small interval of time]. If
\begin{eqnarray*}
p &=& 1 - \delta \theta\\
r &=& 1 + a\theta
\end{eqnarray*}
in the limit when the time-interval $\theta$ is made very small we may write [the number of species at time $t$] $y = e^{(a-\delta)t}$.''
\end{quote}
Today, another important feature of \cite{Y} which is often overlooked is the fact that Yule does not only introduce the pure-birth process but studies a pure-birth process of pure-birth processes, or \textbf{nested} pure-birth process. Thus, even if \cite{Y} does not explicitly address the notion of tree (see Section \ref{subsec:yule-tree}), we can say that if the Yule 1925 model provides the basis for what we call today a Yule tree, it also does for a Yule tree of Yule trees. 

Nested trees, also called `trees within trees' \citep{page1998trees}, have experienced a revival in 21st century evolutionary biology to model a variety of phenomena like  gene trees within the species tree \citep{pamilo1988relationships, maddison1997gene, degnan2009gene, mirarab2021multispecies}, gene family tree within the species tree \citep{rasmussen2012unified, szollosi2015inference} or viral genealogy within the transmission tree \citep{grenfell2004unifying, volz2013viral}... 
Nested trees also pose deep theoretical questions which have triggered the interest of the mathematical community \citep{semplesteel, mehta2016probability, blancas2018trees, blancas2019nested, duchamps2020trees, lambert2020coagulation, mossel2012phylogenetic, mossel2011inference}. It is amusing to see that if the literature about nested trees readily cites \cite{Y} for his paternity of the pure-birth process (or actually of the `Yule tree'), it is never for his study of nested pure-birth processes.
\\ 

Another striking aspect of \cite{Y}'s investigations is to seek to explain empirical patterns (here, distribution of number of species per genus) as resulting from a \textbf{neutral} process, that is, regardless of selective mechanisms. In the following excerpts, Yule even argues in favor of a saltationist view of evolution, before giving his understanding of (mass) extinctions:
\begin{quote}
[p. 22]
``On the Darwinian view that species are continually dying out---that a species rises, flourishes and dies, superseded by the more advantageous form---a species occupying a very small area may be young, but it is equally likely or more likely to be old (a dying
species). On Darwin's own view that the whole body of individuals in a species becomes altered together, the young, species must be found occupying a large area at once, and the species occupying a small area could only be a `dying' species. On the Darwinian view therefore either there need be no relation between Age and Area, or there would be a negative relation, species occupying small areas being on the whole the oldest.\\
Similarly, on the Darwinian view a genus of a few, or of only one, species may be either young or old---a dying genus---and there need be no necessary relation between Age and Size. That species occupying very small areas, and the species of monotypic genera are mainly `relic' forms, is, I gather, the predominant Darwinian view. Dr. \textsc{Willis}'s
conclusions are inconsistent with that view.\\
We are accordingly led directly to the mutational view of evolution that has been held by more than one writer both before Darwin and after : the view that specific differences arise, not cumulatively by the natural selection of slight favourable variations, but at once \textit{per saltum} as `mutations'. On this view a new form must necessarily occupy a small area, and the required correlation between age and area follows at once''.\\

\label{page:cataclysm}
[p. 23]
``Now a `cataclysm' in the sense explained would kill out the whole or a great part of the organic life existing in the region over which it swept. It would necessarily act \emph{differentially}, for some only of all the species in the world would lie within its range, but it would not act \emph{selectively} if the cataclysm was overwhelming and the extermination complete : the species exterminated would be killed out not because of any inherent defects but simply because they had the ill-luck to stand in the path of the cataclysm.''
\end{quote}

 


It would be anachronistic to claim that these ideas foreshadow the advent of neutral theories of molecular evolution \citep{kimura1983} and of biodiversity and biogeography \citep{hubbell2001}. They most surely belong to the stream that remained reluctant to some of Darwin's ideas and survived until the modern synthesis, inspired to Yule by the theories of  J.C. Willis, which culminated in his 1940 book \citep{willis1940}. However, we note that the revival in the 1970's of neutral models of macroevolution led by Woods Hole paleontologists and dealing with the question of distinguishing between deterministic and stochastic causes of macroevolution (see short review in \citealp{mooers1997inferring}), started with stochastic simulations of the Yule tree \citep{raup1973stochastic, gould1977shape}.\\
 
Last, it is wonderful to see the mathematician Yule spend a good third of his revolutionary math paper on confronting his predictions with real \textbf{data}, fitting abundance distributions and estimating parameter values with such enthusiasm:
\begin{quote} 
[p. 27] 
``If we form a chart in which the number of genera of a given size is plotted vertically and the size of the genus horizontally, not to ordinary scales but to logarithmic or ratio scales, that is scales on which numbers that bear equal to each other (like 1, 2, 4, 8, 16) stand at equal distances apart, the resulting points in any actual case run rather irregularly but fairly closely round a straight line, usually up to genera of 30 species or so, sometimes even up to genera of 100 species or more, after which the points fall
rather abruptly away from the line. (...)\\
The numbers of monotypic genera observed and calculated must agree within a decimal point or so owing to the method
of fitting : but I think the reader who studies Tables V to VIII will admit that the agreement between observation and calculation is throughout extraordinarily close. It is in fact better than one has any right to expect. I admit very considerable difficulties of interpretation and would refer to the discussion on pp. 58-62. Here I would only direct attention to the rather large number of primordial genera found in each case (Table IV, line 7, p. 54) : to the comparatively limited range of values of $\tau$ (4.26 to
6.28, ibid, line 5) : and to the comparatively limited range also of the values of $\rho$ (1.188 to 1.925, ibid, line 6). Subject to the admitted difficulties of interpretation, the results of this test, on the one point on which direct comparison can be made with the facts, could hardly be better''.
\end{quote} 
\cite{fisher1943relation} will also use abundance distributions as sources of data, in a work which shares other similarities with \cite{Y}: neutral model, collaboration between a statistician and naturalists, huge audience ($>4000$ citations in 2024).

  




\subsection{Did you say ``Yule tree''?}
\label{subsec:yule-tree}
It is common today to think equivalently of the Yule process and of the Yule tree. The pure-birth process, also called Yule process, counts the number of particles (species, genera) as they accumulate through time, whereas the Yule tree is a representation of the genealogy naturally underlying the Yule process.

A quick text search through \cite{Y} shows that Yule never uses the words `birth' or `division' and that the paper is completely devoid of the notions of tree, genealogy or even lineage. In Yule's words, new species and new genera occur as a consequence of mutations\footnote{It is most likely that the term ``mutation'' here carries its old meaning of a spontaneous large change resulting in the sudden appearance of a new species, before the word took its present meaning of a discrete genetic change, due to (H. de Vries and) T.H. Morgan.
}: species undergo ``specific mutations'' and genera undergo ``generic mutations'' but they are never said to give birth or divide like individual particles would do. However, Yule uses once the word `offspring' and twice the word `generation', all in the same sentence:
\begin{quote}
[p. 40] 
``When work on the frequency distributions of sizes of genera was first begun, considerations of a very rough kind suggested that the limiting form of the distribution for infinite time should approach this logarithmic-linear law. 
The generation of species from species, or genera from genera, seemed closely parallel to the generation of offspring in a given stock in which mortality might be ignored. 
%
(...) The method of approach was obviously exceedingly crude, but it suggested logarithmic plotting of the data''.
\end{quote}
The vagueness of this parallel and its unique appearance in the entire paper suggest that in the ``generation of species from species'', the filiation between the old species and the new species seems to reduce to the fact that the ``specific mutation'' giving rise to the new species can be traced back to (an individual member of?) the old species (the ``stock''). But this trace never takes the form of a species lineage and consequently even less of a species tree:
\textbf{what we call today the \emph{Yule tree} was absent from Yule's original work}. 

Actually, the mathematical research on random genealogies, pioneered independently by Bienaym\'e and by Galton \citep{bienayme1845, heyde1977, galton1874} in works concerned with the extinction of family names, resurfaced\footnote{See also \cite{steffensen1933deux}.} thanks to \cite{otter1949multiplicative}, who introduces a mathematical space of trees (again without citing \citealp{Y}), 
as testified by the following quote of \cite{kendall1949stochastic}:

\begin{quote}
``Of the other new work which has appeared since my paper was written the most important appears to be that of Richard Otter, who has linked up the problems of the growth and extinction of populations with the combinatorial theory of ``trees''. This was being discussed in the pages of the Educational Times only a few years after Galton there proposed his ``surname'' problem, and we may wonder equally at the versatility of the contributors to that remarkable periodical and at the near-century which has had to elapse for these two topics to become reunited.''
\end{quote}

We will precisely seek to understand the dependencies between ages and sizes of different genera as driven by the hidden phylogenetic structure. We will see in Section \ref{sec:rejuvenation} that this introduction of the phylogeny, absent from Yule's work, will allow us to extend his results to models where species and genera may become extinct.

\subsection{What \cite{Y} actually proved}

Here we record the main mathematical results obtained by \cite{Y}. 

We denote by $(Y_t; t\ge 0)$ the pure-birth process with rate $s$ (modeling the number of species in one genus) and by $(N_t; t\ge 0)$ the pure-birth process with rate $g$ (modeling the number of genera).

First, \cite{Y} is interested in the frequency $f_n$ of genera with $n$ species when $N_t$ is constant equal to $N$ ($g=0$, no ``generic mutations''): 
\begin{quote} 
[p. 33]  ``Then, putting aside generic mutations altogether for the present, if we start with $N$ prime species of different genera... (...) The general form of the law is obvious. We have
\begin{equation}
\left.\begin{array}{rcl}
f_1 &=& e^{-st}\\
f_2 &=& e^{-st}(1-e^{-st})\\
f_3 &=& e^{-st}(1-e^{-st})^2\\
\cdots &\cdots&\cdots\cdots\cdots\cdots\cdots\\
f_n &=& e^{-st}(1-e^{-st})^{n-1}\\
\end{array}
\right\}
\label{eqn:yule1}
\end{equation}
That is to say, if $N$ prime genera start together at zero time when they are all monotypic, after time $t$ we will find the numbers that have $1, 2, 3,...$ species given by a geometric series of which the common ratio is $1-e^{-st}$, $s$ being a constant proportional to the chance of a specific mutation occurring in a given time.
(...) 

As regards the increase in the number of genera, the whole process will proceed on precisely the same lines (...) where $g$ is a constant proportional to the chance of a generic mutation (or mutation from genus to genus) occurring in a given time.''
\end{quote} 
In modern terms, we can apply the law of large numbers to the $N$ independent genera, so that $f_n\sim \PP(Y_t=n)$ as $N\to\infty$. As a conclusion, we can say that Yule proved the following result: 
$$
\PP(Y_t=n) = e^{-st}(1- e^{-st})^{n-1}\qquad t\ge 0, n\ge 1,
$$
that he calls ``the frequency distribution of sizes of genera all of the same age''.

Of course, when $g\not=0$, $N$ is also a pure-birth process, so that $\PP_1(N_t=n) = e^{-gt}(1- e^{-gt})^{n-1}$, where the subscript means that $N_0=1$. In particular, recalling the expectation of a geometric distribution, Yule gets
\begin{equation}
\EE_N(N_t) = Ne^{gt}.
\label{eqn:yule2}
\end{equation}

Then, Yule seeks to relax the assumption that $g=0$ and derive the frequency of genera with $n$ species in the presence of phylogenetic correlations between the extant genera. Since the diversification processes of different genera are independent conditional on the ages of the genera, Yule naturally uses the disintegration over these ages:
\begin{quote} 
[p. 37]  ``We first require to know how many out of the totality of genera existing at any given time, say $T$, are of any assigned age $x$. From (8) [here, \eqref{eqn:yule2}] the total number of genera at time $t$ is $Ne^{gt}$. The number coming into existence during the interval $\pm\frac12 dt$ round time $t$ is therefore $Nge^{gt}dt$, and the number of age $x$ at time $T$ is
$$
Nge^{g(T-x)}dx.
$$
The \textit{proportion} aged $x$ at time $T$ is therefore $ge^{-gx}dx$. Note that these are the \textit{derived} genera only, ignoring the prime genera with which we started ; but, as stated at the end of Section I, since we are going to take time as infinite the number of derived genera will be infinitely great as compared with the number of primordial genera, and the latter may legitimately be ignored during the present stage of the work.

We have now got to take the series (5) [here \eqref{eqn:yule1}], writing $x$ for $t$ throughout, term by term, multiply each term by $ge^{-gx} dx$ and integrate from zero to infinity.'' 
 \end{quote} 
 Let us expand a little bit this reasoning. Let $N_t(dx)$ (resp. $N_t(n, dx)$) denote the number of genera at time $t$ with age $\in (x, x+ dx)$ (resp. with $n$ species and age $\in (x, x+dx)$).
Let $N_t(n)$ (resp. $N_t^*(n)$) denote the number of genera at time $t$ with $n$ species, including (resp. to the exception of) the primordial genus. What Yule actually proves is
$$
\EE_N(N_T(dx)) = Nge^{g(T-x)}dx.
$$
Note that the factor $N$ is useless here because the $N$ initial genera have identically distributed descendances, so we will take $N=1$ from now on, without loss of generality.
Then he implicitly uses the fact that $\EE_1(N_T(n, dt)) = ge^{g(T-t)}\PP(Y_t=n)\,dt$, and despite the fact that he never writes it explicitly,
\begin{equation}
\EE_1(N_T^*(n)) = \int_0^T ge^{g(T-t)}\PP(Y_t=n)\,dt,
\label{eqn:yule3}
\end{equation}
and
\begin{equation}
\EE_1(N_T(n)) = \int_0^T ge^{g(T-t)}\PP(Y_t=n)\,dt + \PP(Y_T=n),
\label{eqn:yule3bis}
\end{equation}
where the last term is the contribution of the `primordial' genus. 
Now define
$$
f_n^Y(T):=\frac{\EE_1(N^*_T(n))}{\EE_1(N_T)}\quad\mbox{ and }\quad f_n^Y:=\lim_{T\to\infty}f_n^Y(T),
$$
that \cite{Y} terms the ``frequency distribution for sizes of derived genera'', (which he still denotes $f_n$, causing a conflict of notation).
Then we get 
\begin{equation}
f_n^Y(T)= \int_0^T ge^{-gt}\PP(Y_t=n)\,dt \quad\mbox{ and }\quad f_n^Y= \int_0^\infty ge^{-gt}\PP(Y_t=n)\,dt.
\label{eqn:yule4}
\end{equation}
It is precisely this quantity that Yule computes further in the paper, although never writing it down under its integral form.
\begin{rem}
As $T$ gets large,
\begin{equation}
\frac{\EE_1(N^*_T(n))}{\EE_1(N_T)} \sim\frac{\EE_1(N_T(n))}{\EE_1(N_T)} \sim \int_0^\infty ge^{-gt}\PP(Y_t=n)\,dt.
\label{eqn:yule5}
\end{equation}
Thanks to the works of the Swedish branching process school on exponentially growing processes counted with a characteristic  \citep{Ner81, JN84a, JN84b}, we also know that we can drop the expectations in the previous convergence. More specifically, as $T$ gets large, the following convergence holds almost surely:
\begin{equation}
\frac{N_T(n)}{N_T} \sim \int_0^\infty ge^{-gt}\PP(Y_t=n)\,dt.
\label{eqn:jagers}
\end{equation}
This convergence result was proven, using the same technique, by \cite{holmgren2017fringe}, who also refer to \citet[\S 6.10.3]{Johnson}, see (their) Example B.11 of (their) Theorem 5.14.
\end{rem}

Note that we can write 
\begin{equation}
\label{eqn:fnY}
f_n^Y(T)=\PP (Y_D=n, D<T)
\end{equation}
where $D$ is independent of the pure-birth process $Y$ and follows the exponential distribution with parameter $g$.
It is tempting  to believe that $(D,Y_D)$ is the law of the age and size of a randomly chosen genus but for now, \eqref{eqn:fnY} merely provides an expression for $f_n^Y(T)$, defined by \cite{Y} as a ratio of expectations of quantities defined at the population (genus) level. We will thus seek to answer the following natural questions:
\begin{itemize}
\item[(Q1)] Is $(D,Y_D)$ the law of age and size of a randomly chosen genus?
\item[(Q2)] If yes, what is the structure of correlations of ages and sizes between the different genera? 
\item[(Q3)] For more general species diversification processes, including in particular extinctions, can we answer the previous questions? If yes, how do sizes and ages of genera covary with their phylogeny? 
\item[(Q4)] Can we display necessary and/or sufficient conditions for such nested processes, and the associated urn schemes, to produce size distributions with power-law tail?
\label{rem:q}
\end{itemize}

\section{Rejuvenation of \cite{Y}: More ages and sizes, more trees and urns}
\label{sec:rejuvenation}

\subsection{Framework}
We try to adopt a notation which sticks as closely as possible to the notation of \cite{Y} but also leaves room for the extensions we wish to present. In particular, we now allow for species and for genera to become extinct. 
\begin{itemize}
\item We denote by \textbf{$N_t$ the number of genera extant} at time $t$ and we stop the process $(N_t;t\ge 0)$ at time $T$ called \emph{present time}. 

We call $(N_t;0\le t\le T)$ the \textbf{generation process};
\item Each genus $i$ born at time $t_i$ contains a random \textbf{number $Z_t^{(i)}$ of species} at time $t_i+t$. 

We call the within-genus dynamics $(Z_t^{(i)};t\ge 0)$ the \textbf{diversification processes}.
\end{itemize}
Now here are our generic assumptions:

\begin{itemize}
\item[(i)] The diversification processes are \textbf{independent and identically distributed} (i.i.d.).

They have the same transition probabilities and initial value as a time-continuous, time-homogeneous Markov process $(Z_t;t\ge0)$;
\item[(i')] A genus containing no species is considered \textbf{extinct}. 

An extinct genus cannot gain new species (i.e., 0 is an absorbing state for the Markov process $Z$).
\item[(ii)] Each \textbf{extant genus gives birth independently at rate $g$} to daughter genera, one at a time, independently of the diversification process it harbors, provided it is non-extinct.
\item[(iii)] At time $T$, we \textbf{sample independently each extant genus with probability $f$} and we condition the size of the sample to be nonzero.
\end{itemize}
This framework and these assumptions are common to \cite{Y} and to the present paper, but \cite{Y} adopts the following restrictions:
\begin{itemize}
\item[(Y1)] The diversification process \textbf{$(Z_t;t\ge 0)$ is a pure-birth process with rate $s$} started at $Z_0=1$. 

As a consequence, no genus can become extinct so Assumption (i') actually is irrelevant and because of (ii):
\item[(Y2)] The generation process \textbf{$(N_t;t\ge 0)$ is a pure-birth process with rate $g$}.
\item[(Y3)] $f=1$.
\end{itemize}

Let us already make a few remarks about these assumptions.
\begin{rem}
\label{rem:orientation}
It is implicit that at each birth event of a new genus, we distinguish the \textbf{mother genus} and the \textbf{daughter genus}: the number of species in the mother genus is not perturbed by the birth event, while the number of species in the daughter is equal to $Z_0$ upon birth.
\end{rem}

\begin{rem}
Let $L$ denote the typical genus \textbf{lifetime}, i.e.,
\begin{equation}
L:=\inf\{t>0: Z_t=0\},
\end{equation}
where $L:=+\infty$ in case $Z$ does not hit 0.
Then the generation process, which counts the number of genera through time, is not Markovian unless $L$ is exponential (or infinite as under Assumption (Y1)). However, because of Assumptions (i) and (ii), it is a branching process, called a \textbf{Crump-Mode-Jagers process}. 
\end{rem}

\begin{rem} According to (ii), the `generic births' occur at constant rate $g$ per genus and not at a rate which depends, e.g., linearly with the number of species it contains, as Simon would later assume (see Section \ref{sec:urns}) as seems more natural and as the terminology of generic `mutation' could lead to think. The idea of Yule is that these `mutations' occur in individuals and that the number of individuals in a genus remains roughly constant:
\begin{quote} 
[p.24] ``At first, as the species spreads, the number of individuals must tend to increase. But over the very long periods which have to be considered there must be a countervailing tendency to ultimate decrease in the number of individuals, owing to the increase in the number of species. The area available being limited, the tendency, as it seems to me, must be towards greater
and greater numbers of species and fewer individuals in each. (...)

The possible effect of size of genus (number of species in the genus) on the chance of a generic mutation is also ignored.
This assumption may or may not be correct, but was deliberate. The generic characters are regarded as representing a main position of stability, and the chance of occurrence of a transfer from one main position of stability to another is regarded as independent of the number of minor positions of stability (species) which may have been taken up within the main position (genus).''
\end{quote}

\end{rem}

\begin{figure}[h] 
\begin{center} 
\includegraphics[width=.8\textwidth]{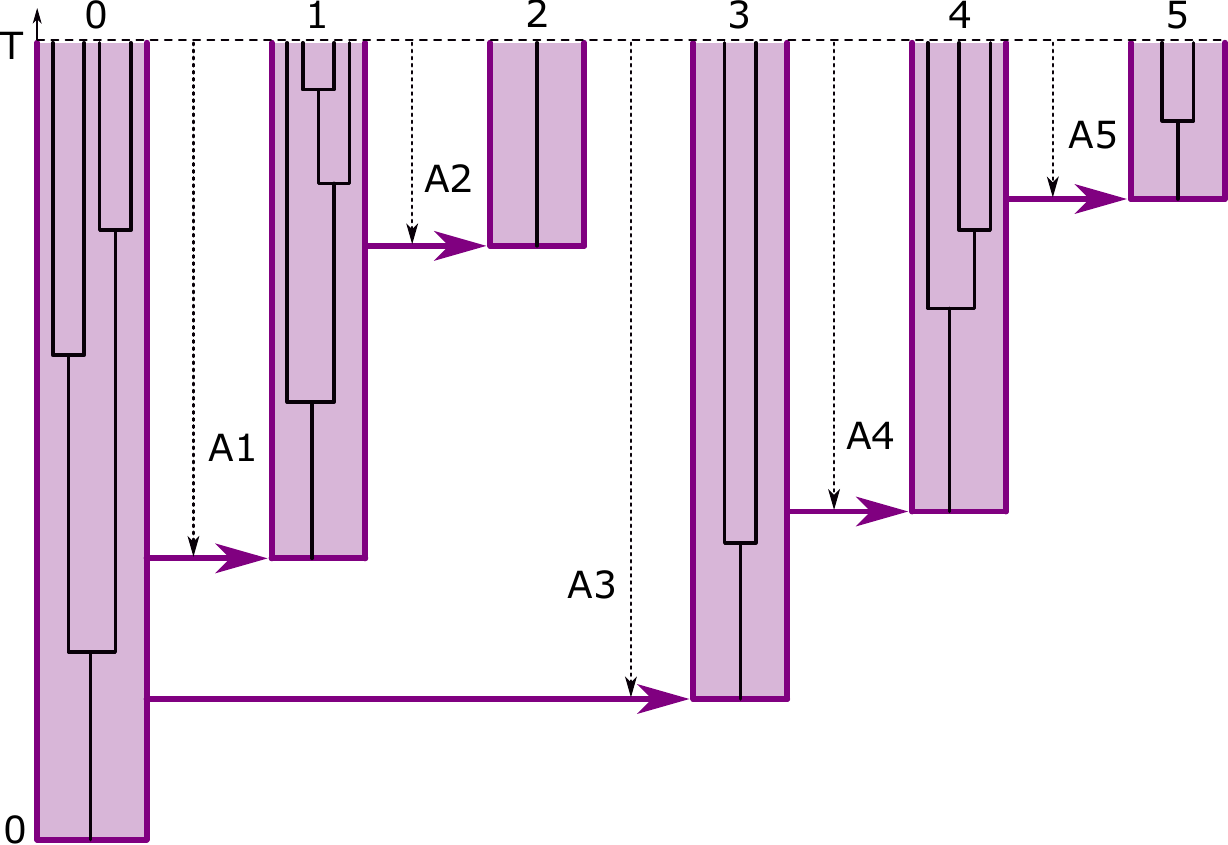} 
\end{center}
 \caption{\textbf{Plane representation of a Yule tree of Yule trees.} Drawing daughter genera to the right of their mother genus and forbidding horizontal birth arrows to intersect vertical branches induces an orientation and a labeling from left to right of genera, starting at $0$ (primordial genus) and ending at $N_T-1$. Vertical dotted arrows show the ages $A_1,\ldots, A_{N_T-1}$ of the derived genera.} 
 \label{fig:yiy-cpp}
 \end{figure}

\subsection{Rejuvenation I: Joint law of ages and sizes of derived genera in Yule 1925 model}

\paragraph{Initial condition.} \cite{Y} often assumes that the clade is initially composed of $N$ genera, because observables like the number of species in a genus are described in terms of frequencies taken over a population of $N$ i.i.d processes initiated by $N$ ``primordial'' genera (as opposed to ``derived''). Here we will adopt a more probabilistic view and express such quantities as probabilities, in such a way that the number of initial genera is irrelevant. Hence we assume $N_0=1$. We will also assume $Z_0=1$ as Yule does.

\paragraph{Orientation and labeling.} Recall from Remark \ref{rem:orientation} that birth events in the `Yule tree' are asymmetric, allowing us to distinguish a mother and a daughter genus. Drawing the genus tree in the plane by always letting the \textbf{mother-to-daughter arrows go from left to right} and forbid arrows and branches to intersect as shown in Figure \ref{fig:yiy-cpp}, induces an \textbf{orientation} in the tree and especially in the genera extant at present time $T$. We can then label extant genera from left to right, starting at $i=0$ which is the label of the `primordial genus', up to $i=N_T-1$.\\

Now set 
$$
A_i:= \mbox{ \bf age of genus } i \quad\mbox{ and } \quad S_i:=\mbox{ \bf size of genus } i , 
$$
where the size of a genus is the number of species it harbors at present time. \\

Now assume that \textbf{[$(Y_1), (Y_2), (Y_3)$] are in force} as in \cite{Y}. 
In particular, $A_0=T$ and $S_0$ has the law of $Y_T$. 

\begin{thm}\label{thm:yule1} 
The pairs $((A_i, S_i);1\le i \le N_T-1)$ of ages and sizes of derived genera extant at time $T$, form a sequence of \textbf{independent} random pairs all distributed as a pair $(A,S)$, killed at the first value of its first component larger than $T$. The pair $(A,S)$ has the following properties:
\begin{itemize}
\item The age $A$ follows the exponential distribution with parameter $g$
\item
For any $a>0$ and $k\ge 1$, 
$$
\PP(S=k\,|\,A=a)=\PP(Y_a=k).
$$
\end{itemize}
As a consequence,
\begin{equation}
\PP(S=k, A\in da)= ge^{-(g+s)a}(1- e^{-sa})^{k-1}\qquad a>0, k\ge 1.
\end{equation}
\end{thm}
\begin{rem}
\label{rem:yule1} 
From the first sentence of Theorem \ref{thm:yule1} we recover the fact that $N_T$ follows the geometric distribution with success probability $\PP(A>T)=e^{-gT}$.
\end{rem}
Thus, we have answered Questions (Q1) and (Q2) of p.\pageref{rem:q}: (Answer to Q1) The joint law of age and size of each and every genus is indeed given by $(D, Y_D)$ and (Answer to Q2) these pairs are independent. Let us proceed with the more general setting to address Question (Q3) about extensions of Theorem \ref{thm:yule1} to general diversification processes, including in particular extinctions, and to the phylogeny.

\subsection{Rejuvenation II: Extension to general diversification processes}

Here, the restrictions \textbf{[$(Y_1), (Y_2), (Y_3)$] are not in force} any longer. 
In particular, each genus extant at time $T$ is independently sampled with probability $f$. We call $N_T^f$ the \textbf{number of sampled genera}, and we work conditional on $N_T^f\not=0$.

Recall that $L$ denotes the \textbf{lifetime} of a genus and notice that the number of offspring genera per genus has mean $g\EE (L)$. Then the generation process is a supercritical branching process as soon as
$$
g\EE (L)>1.
$$
In this case, the so-called \textbf{Malthusian parameter} $\eta$ of the generation process is defined as the only positive root of the so-called \textbf{characteristic exponent} $\psi$ defined as
\begin{equation}
\label{eqn:psi}
\psi(x) = x-g +g\EE \left(e^{-xL}\right)\qquad x> 0.
\end{equation}
If $g\EE (L)\le 1$, we set $\eta=0$. From now on, we will assume that $g\EE (L)\ge 1$.
Also notice that 
$$
\psi(0+):=\lim_{x\to 0, x>0} \psi(x) = -g\PP (L=\infty).
$$
 
\begin{figure}[h] 
\begin{center} 
\includegraphics[width=.8\textwidth]{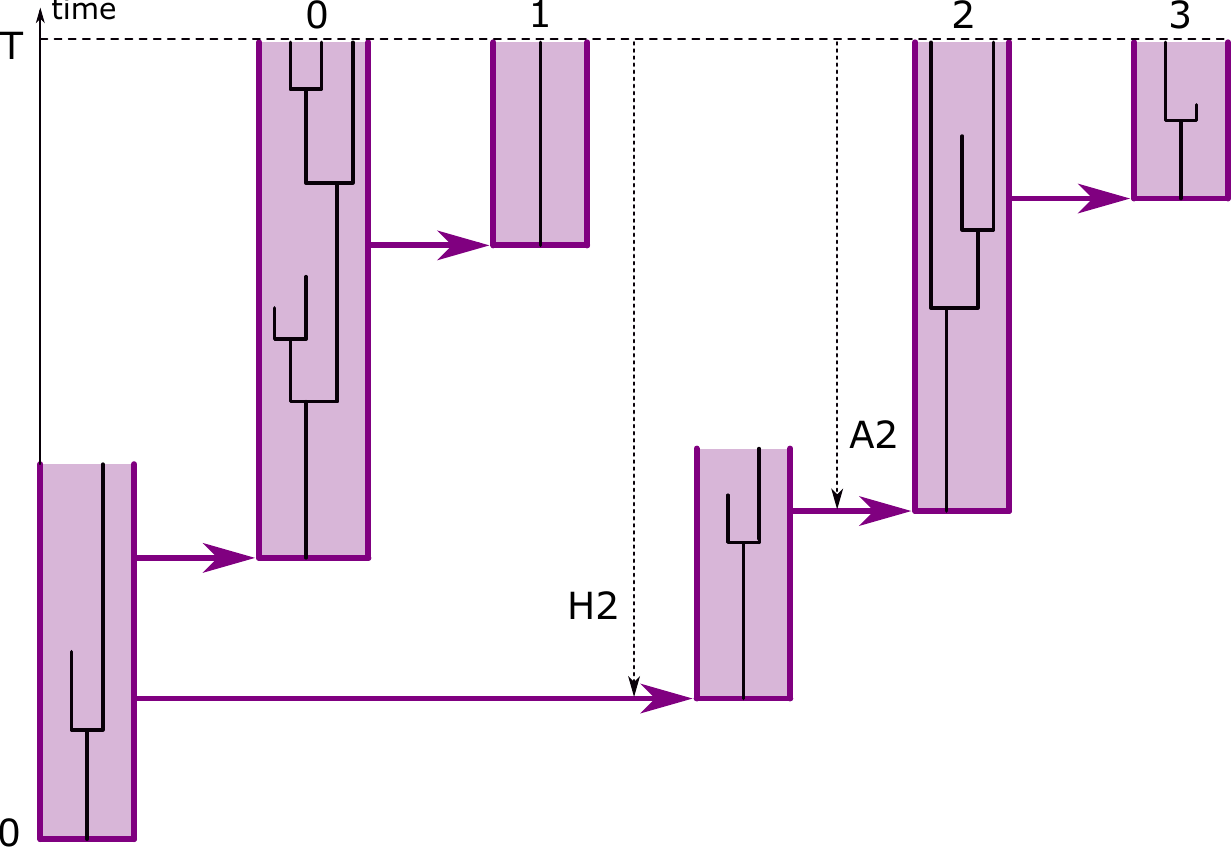} 
\end{center}
 \caption{\textbf{Tree of trees in the presence of extinctions.} The orientation of the tree follows the same rules as in Figure \ref{fig:yiy-cpp} and induces a labeling on genera extant at $T$ (here $f=1$: full sampling). 
 Vertical dotted arrows show the age $A_2$ of genus 2 and the coalescence time $H_2$ between genera 1 and 2.} 
 \label{fig:bdiy}
 \end{figure}

We stick to the orientation defined in the previous section and shown in Figure \ref{fig:bdiy}, so we can label sampled genera from left to right, starting at $i=0$ (that is not necessarily the primordial genus, which can be extant but not sampled, or extinct) up to $i=N_T^f-1$. 

We also stick to the notation $A_i$ (resp. $S_i$) for the age (resp. size) of the $i$-th sampled genus and we additionally define
$$
H_i:= \mbox{ \bf coalescence time between genus } i-1 \mbox{ \bf and genus } i, 
$$
where the labels refer to sampled genera and the time is measured backward from time $T$ (so that the coalescence time is actually a `coalescence depth'). Notice that we always have $H_i\ge A_i$ (with a.s. equality only when $f=1$ and in the absence of extinction). 
\\

Now observe that the number of species in a genus is what is called a \textbf{non-heritable trait}, i.e., is a trait which can evolve randomly through the lifetime of the genus, independently of times when the genus gives birth, and whose initial value at the birth of the genus is drawn independently from a given distribution (here $Z_0=1$). 
\\

Thus, we can rely on \textbf{coalescent point process theory} \citep{AP05, lambert2010contour, lambert2013birth} which basically ensures that a certain traversal of the tree from left-to-right has remarkable renewal properties (see Appendix, Section \ref{subsec:joint}), whenever the tree is produced by a binary branching process where the birth rate depends at most on time and the death rate depends at most on time and a non-heritable trait. 

Let $W$ be the unique increasing solution with initial condition $W(0)=1$ to 
\begin{equation}
\label{eqn:integro-diff}
W'(t) = g\left(W(t) - \int_0^t W(s) \, v(t-s)\, ds\right)\qquad t\ge 0,
\end{equation}
where $v$ is the density of the lifetime $L$ of a genus. The function $W$ is given implicitly by its Laplace transform
$$
\int_0^\infty W(y) \,e^{-xy}\,dy = \frac1{\psi(x)}\qquad x>\eta,
$$
where $\psi$ is defined in \eqref{eqn:psi}.
Also set
$$
W_f(t) := 1-f +fW(t)\qquad t\ge 0.
$$
We can use these ingredients to state our main result, which is proved in Appendix, Section \ref{subsec:joint}. 
\begin{thm}\label{thm:yule2} 
The triples $((H_i, A_i, S_i);1\le i \le N_T^f)$ of coalescence times, ages and sizes of genera sampled at time $T$, form a sequence of \textbf{independent} random triples all distributed as a triple $(H, A,S)$, killed at the first value of its first component larger than $T$. The triple $(H,A,S)$ has the following properties:
\begin{itemize}
\item
$H$ and $S$ are independent conditional on $A$, whose law is given by
\begin{equation}
\label{eqn:lawA}
\PP(A\in da) = ge^{-\eta a} \PP(L>a)\, da.
\end{equation}
\item
The joint law of $(H,A)$ is given by
\begin{equation}
\label{eqn:density-age-1}
\PP (H<h, A\in da)= \frac{fW(h-a)}{W_f(h)}g\PP(L>a)\, da\qquad 0<a\le h.
\end{equation}
Letting $h\downarrow a$ yields
\begin{equation}
\label{eqn:a=h-1}
\PP (H=A\in da)= \frac{f}{W_f(a)}g\PP(L>a)\, da.
\end{equation}
\item
The joint law of $(A, S)$ is given by
$$
\PP(S=k\,|\,A=a)=\PP(Z_a=k\,|\, Z_a\not=0)\qquad k\ge 1, a>0,
$$
so that 
\begin{equation}
\label{eqn:jointSA}
\PP(S=k,A\in da) 
 =g\PP(Z_a=k)e^{-\eta a}da.
\end{equation}
\end{itemize}
\end{thm}
\begin{rem}
\label{rem:geom}
Similarly as in Remark \ref{rem:yule1}, we note that the number $N_T^f$ of genera sampled at $T$ follows the geometric distribution with success probability $\PP(H>T)$, whose expression depends on $g$ but also on the law of $L$, and is actually equal to $1/W_f(T)$ (see Section \ref{subsec:joint}).
\end{rem}
\begin{rem} Recall from \eqref{eqn:psi} that $\psi(x) = x-g +g\int_0^\infty e^{-xt}v(t)\, dt$ and that $\psi(\eta)=0$. An integration by parts yields $\psi(x) = x-gx\int_0^\infty e^{-xt}\PP(L>t)\, dt$, which evaluated at $\eta$ gives $\int_0^\infty ge^{-\eta t}\PP(L>t)\, dt=1$, showing that the law \eqref{eqn:lawA} of $A$ indeed integrates to 1.
\end{rem}

\begin{rem}
Reasoning similarly as Yule and relying on the fact that in the presence of extinctions (but provided $g\EE (L)>1$), genera grow exponentially in number with rate $0<\eta<g$, we could show that the fraction of derived (and sampled) genera with size $k$ and age $\in da$ is $g\PP(Z_a=k)e^{-\eta a}da$. It would be tempting to say that ages of genera are now exponential with parameter $\eta$, but we know from Theorem \ref{thm:yule2} that this is not true.
\end{rem}

\begin{rem}
Coalescent point process theory actually ensures that the previous statement still holds if we assume that an additional cause of genus extinction (than the extinction of the diversification process it harbors, see (i')) is a series of \textit{mass extinction events}: there are fixed times $0<s_1< s_2<\cdots< s_\ell<T$, called mass extinction times, such that for each $1\le i\le \ell$, each extant genus at time $s_i$ is independently killed with some given probability $q_i$ (with all the species it contains at this time). 
Although \cite{Y} did not implement this possibility, he mentions `cataclysmic killing' as the main, if not only in his view, cause of genus extinction (see quote of \citealt[p.23]{Y}, reproduced p.\pageref{page:cataclysm} of the present article).
\end{rem}

Recall Question (Q3) of page \pageref{rem:q}: ``For more general species diversification processes, including in particular extinctions, can we answer questions Q1 and Q2? If yes, how do sizes and ages of genera covary with their phylogeny?'' 
Theorem \ref{thm:yule2} gives the answer to this question when the species diversification process is any time-homogeneous, integer-valued Markov process $Z$ for which $0$ is absorbing and when the generation process is as in \cite{Y} except that genera containing no species are extinct and cannot give birth to new genera.

We now seek to answer Question (Q4): ``Can we display necessary and/or sufficient conditions for such nested processes, and the associated urn schemes, to produce size distributions with power-law tail?''
To this aim, we study two special cases of Theorem \ref{thm:yule2}: 
\begin{itemize}
\item  \textbf{Case 1:} $Z$ is a time-homogeneous, \textbf{linear birth--death process} with per capita birth rate $\lambda$ and per capita death rate $\mu$;
\item \textbf{Case 2:} $Z$ is a time-homogeneous, \textbf{general pure-birth process} which (only) jumps by $+1$, at rate $\lambda_k$ when in state $k$.
\end{itemize}
Cases 1 and 2 precisely intersect in the Yule case ($\lambda=s$ and $\mu=0$ in Case 1, $\lambda_k=ks$ in Case 2).
We now state our results in each of these two cases, in the form of Proposition \ref{prop:case1} and Proposition \ref{prop:case2}, to be proved in Appendix, Section \ref{app:prop}.

\paragraph{Case 1.} When $\lambda\ge \mu$, $\EE(L)=\infty$ so the generation process is supercritical whatever $g>0$. When $\lambda<\mu$ it is well-known (see e.g., \citealp[p. 78]{lambert2008population}) that
$$
\EE(L) = \frac{\ln (\mu)-\ln (\mu-\lambda)}{\lambda}.
$$
In this case, we set
$$
g^\star:= \frac{1}{\EE(L) }=\frac{\lambda}{\ln (\mu)-\ln (\mu-\lambda)}
$$
so that the generation process is critical if $g=g^\star$ (in which case $\eta=0$) and supercritical if $g>g^\star$ (in which case $\eta>0$). We thus assume that $\lambda\ge \mu$ and otherwise that $g\ge g^\star$.

\begin{prop}
\label{prop:case1}
Assume that $Z$ is a linear birth-death process with birth rate $\lambda$ and death rate $\mu$.
The law of the number $S$ of species per genus is given by
$$
\PP(S=n)=\frac g\lambda\int_0^\infty e^{-\eta t}\frac{F'(t)}{F(t)^2}\left(1-\frac1{F(t)}\right)^{n-1}dt, 
$$  
where $F(t) =1+\lambda (e^{rt}-1)/r$ when $r:=\lambda-\mu\not=0$ and $F(t) =1+\lambda t$ if $r=0$. 

If $\lambda<\mu$ then we have the following inequality if $g>g^\star$
$$
\PP(S=n)< \frac{g}{\lambda}\frac{(\lambda/\mu)^n}{n},
$$ 
and if $g=g^\star$, the genus size distribution is given by Fisher's log-series:
$$
\PP(S=n)=\frac{g^\star}{\lambda}\frac{(\lambda/\mu)^n}{n}\qquad n\ge 1.
$$ 
If $\lambda=\mu$, then $S$ has a heavy, but not power-law, tail. More specifically, 
\begin{equation}
\label{eqn:lap}
\lim_{n\to\infty}\ln\big(\PP(S=n)\big)/\sqrt n=-2\sqrt{\eta/\lambda} .
\end{equation}
If $\lambda> \mu$, then $S$ has a power-law tail, in the sense that 
\begin{equation}
\label{eqn:log-log}
\lim_{n\to\infty}\ln\big(\PP(S=n)\big)/\ln n= -\alpha-1,
\end{equation}
where $\alpha = \eta/r$, which is the ratio of the exponential growth rates of the generation process and of the diversification process (similarly as $\alpha=g/s$ when $\mu=0$).

\end{prop}

\paragraph{Case 2.} Here, $Z$ is non-decreasing so that $\PP(L=+\infty)=1$ and $\eta=g$.
\begin{prop}
\label{prop:case2}
Assume that $Z$ is a pure-birth process with jump rate $\lambda_k$ from $k$ to $k+1$. Then
\begin{equation}
\label{eqn:case2-eq1}
\PP (S=n)=\frac{g}{g+\lambda_n}\prod_{k=1}^{n-1} \frac{\lambda_k}{g+\lambda_k}
\end{equation}
and
\begin{equation}
\label{eqn:case2-eq2}
\EE(A\,|\, S=n)  =  \sum_{k=1}^n \frac{1}{g+\lambda_k}.
\end{equation}
Moreover, 
\begin{itemize}
\item\ [non-preferential attachment] If $(\lambda_n)$ is bounded, then $S$ has a rapidly decaying tail;
\item\ [sub-preferential attachment] If $\lim_{n\to\infty}\ln(\lambda_n)/\ln(n) \in (0,1)$, then $S$ has a heavy, but no power-law, tail; 
\item\ [super-preferential attachment] If $\lim_{n\to\infty}\ln(\lambda_n)/\ln(n)>1$, then there is one genus which size blows up in finite time and therefore represents a proportion going to 1 of all species;
\item\ [(asymptotically linear) preferential attachment] If $\lim_{n\to\infty}n^{-1}\lambda_n= s>0$, then $S$ has a power-law tail in the sense that 
\begin{equation}
\lim_{n\to\infty}\ln\big(\PP(S=n)\big)/\ln n= -\alpha-1,
\end{equation}
where $\alpha = g/s$, as in the Yule case where $\lambda_n=sn$.  
\end{itemize}

\end{prop}

\begin{rem}
The previous proposition states in essence that the power law tail of $S$ is only obtained when $\lim_{n\to\infty}\lambda_n/n =s$ is finite and nonzero. In the case of sub-preferential attachment, $S$ has no heavy tail and in the case of super-preferential attachment, a small advantage at the onset of the process tends to snowball over time so that a `winner takes all'. This translates in the discrete urn scheme into the event that one of the genera hosts a proportion of species that tends to 1 with the number of species.
\end{rem}

\subsection{Rejuvenation III: Urn schemes with ``preferential attachment''}
\label{sec:urns}
Here, we recall three well-known 
 urn schemes where a growing number of balls is partitioned into a growing number of urns (or equivalently, into a growing number of colors). Each of these three processes starts with one urn containing one ball and then one ball is sequentially added, one at a time, either into an existing urn or into a new, empty urn. \textbf{When there are $n$ balls partitioned into $k$ urns, with probability $m(k,n)/(m(k,n)+sn)$, the ball is added to a new, empty urn and with probability $sn/(m(k,n)+sn)$ it is added to an existing urn, picked in proportion to the number of balls it contains}. The difference between the three processes is in the choice of $m(k,n)$. We now describe and give a possible continuous-time embedding \citep{athreya1968embedding} for the three schemes:
\begin{itemize}
\item \textbf{[Hoppe urn scheme]} \cite{hoppe1984polya, hoppe1987sampling} Here, $m(k,n)=\theta$. This process is the embedded chain of a \textbf{pure-birth process with immigration}, where new genera (urns) immigrate at rate $\theta$ and species (balls) give birth independently at rate $s$. This urn scheme is also called the Chinese restaurant process (where urns are called tables and balls are called customers) \citep{pitman} and gives rise to Ewens' sampling formula \citep{Ewe72}, Fisher's log-series \citep{fisher1943relation} and the Poisson--Dirichlet distribution \citep{DT86, tavare1987birth}.  See also \cite{Tav24} in same issue.
\item \textbf{[Yule urn scheme]} \cite{Y}  Here, $m(k,n)=gk$. This process is the embedded chain of the Yule model, where genera (urns) give birth independently at rate $g$ and species (balls) give birth independently at rate $s$.
\item \textbf{[Simon urn scheme]} \cite{simon1955class} Here, $m(k,n)=\sigma n$. This process is the embedded chain of a \textbf{pure-birth process with mutations}, where species (balls) divide independently at rate $b=s+\sigma$ and at each birth event, the daughter species belongs to a brand new genus with probability $\beta:=\sigma/(s+\sigma)$ (mutation probability) or to the same genus as its mother with probability $s/(s+\sigma)$. The total number of mutant descendants when there are $n$ balls and $\sigma=a/n$ converges as $n$ gets large to the well-known Luria-Delbr\"uck distribution with parameter $a$ \citep{luria1943mutations, murray2016salvador}. 
\end{itemize}
In other words, these three urn schemes can be seen as the discretization of a pure-birth process of species diversification where new genera arise at a rate which is : \textbf{constant (Hoppe); proportional to the number of existing genera (Yule); proportional to the number of existing species (Simon)}. The frequency distributions emerging in the first and third schemes have been studied (and generalized to non-Markovian diversification processes) by \cite{lambert2011species}.

In the continuous version of each scheme, the number of species in a given genus increases like a pure-birth process with rate $s$. 
Therefore any given genus with age $a$ contains a number of species which is geometric with success probability $e^{-sa}$, regardless of the scheme. As a consequence, differences between the genus size distributions of the different schemes are exclusively driven by differences in genus age distributions: the heavier the tail of the age distribution, the heavier the tail of the size distribution. 

In the Hoppe case (pure-birth with immigration at rate $\theta$), the number of genera (started at 0 at time 0) at time $T$ is a Poisson r.v. $N_T$ with parameter $\theta T$ and conditional on $N_T$, genus ages are independent and uniformly distributed in $(0,T)$. As a consequence, 
$$
f_n^H(T)=\frac 1T\int_0^T da\, e^{-sa}(1-e^{-sa})^{n-1} = \frac{(1-e^{-sT})^n}{sTn}.
$$
 In the Simon case (pure-birth with mutations at birth with probability $\beta$), the number of genera (started at 1 at time 0) at time $T$ is a mixed binomial r.v. $N_T$ with parameter $\beta$ and total number of species which is geometric with success probability $e^{-bT}$. As a consequence, $N_T$ is a geometric r.v. with success probability $e^{-bT}/(e^{-bT}+\beta(1-e^{-bT})))$.
 Now genus ages are independent and follow the exponential distribution with parameter $b$ (because mutation events and node depths are independent) so we have $f_n^S=f_n^Y$, replacing $g$ with $b$ in Equations \eqref{eqn:integral-form}, \eqref{eqn:yule-simon} and \eqref{eqn:power-law}. 
 
 The only difference is that $b=s+\sigma$, so that $b>s$ in Simon's setting, whereas in Yule's setting, $g$ can be smaller than $s$ and is even thought of this way because genera are supposed to renew more slowly than species diversify. As a result, both distributions decay as a power-law with exponent $-1-\alpha$, where $\alpha=b/s>1$ in Simon's setting and in Yule's setting $\alpha=g/s<1$.
Then the distribution $(f_n^S)$
has finite expectation 
$$
\sum_{n\ge 1}n f_n^S =  \int_0^\infty da\, be^{-ba}e^{-sa}\sum_{n\ge 1}n(1-e^{-sa})^{n-1}=  \int_0^\infty da\, be^{-ba}e^{sa}=\frac{b}{b-s} = \frac{\alpha}{\alpha -1},
$$
as opposed to Yule's distribution which has infinite expectation if one assumes $g<s$.
 \begin{rem}
The infinite expectation of $(f_n^Y)$ was commented by \cite{Y}:
\begin{quote} 
[p. 38] 
``The frequency distribution given by the terms of (12) \emph{[here \eqref{eqn:yule-simon}]} is therefore one of those paradoxical distributions in which, though the median, etc., are finite, the mean is infinite. This is, of course, as it should be, for on our assumptions the mean size of a genus after infinite time must itself be infinite.''
\end{quote} 
Yule's conclusion is erroneous. When $\alpha <1$, it is true that most species are to be found in genera founded at times $O(1)$ in particular the primordial genus, which indeed have infinite age in the limit $T\to\infty$; but the sizes of these genera are themselves infinite, not only their mean. Second, the very genera whose law is given by $(f_n^Y)$ have a finite age $A$ with finite expectation $1/g$. Third, the fact that the mean of $(f_n^Y)$ is infinite cannot be structural to the fact that ages are large since this mean is finite when $\alpha >1$ (of course, then $g$ is larger and the mean age $1/g$ is smaller).
\end{rem}
We now consider two extensions of the Yule urn scheme, which are the embedded Markov chains associated with the two applications of the previous section (Case 1 and Case 2). In Case 1, we set $s=\lambda+\mu$. In Case 2, we call $\lambda_k$ the \textbf{weight of an urn} containing $k$ balls. 
 When there are $n$ balls partitioned into $k$ urns,
\begin{itemize}
\item  Case 1: with probability $gk/(gk+ sn)$ a new urn containing one ball is added, with probability $sn/(gk+sn)$ an existing urn is picked in proportion to the number of balls it contains, and then either a new ball is added to the urn (with probability $\lambda/s$) or \textbf{a ball is removed from the urn (with probability $\mu/s$)};
\item Case 2: with probability $gk/(gk+ \Lambda_n)$, where $\Lambda_n$ is the sum of all urn weights, a new urn containing one ball is added, and with probability $\Lambda_n/(gk+\Lambda_n)$ a ball is added to an existing urn \textbf{picked in proportion to its weight} (as opposed to its size).
\end{itemize}
The following informal statement is a corollary to Propositions \ref{prop:case1} and \ref{prop:case2}.
\begin{prop}
\label{prop:cor}
For $i=1,2$, let $(f_n^{(i)})$ be the frequency distribution of urn sizes (i.e., numbers of balls) in Case i. Then
\begin{itemize}
\item $(f_n^{(1)})$ has a \textbf{power law tail iff} $\lambda>\mu$ (more ball additions than ball extractions on average);
\item $(f_n^{(2)})$ has a \textbf{power law tail iff} $\lambda_n$ is asymptotically linear in $n$ (case of classical preferential attachment, where urns are picked in proportion to their size). 
\end{itemize}
\end{prop}
This provides an answer to Question (Q4) on conditions for urn schemes to produce size distributions with power-law tail. To paraphrase the statements in Propositions \ref{prop:case2} and \ref{prop:cor}:
\begin{itemize}
\item If we modify the Yule urn scheme so as to \textbf{pick urns in proportion to a power $\kappa$ of their size}, then \textbf{super-preferential attachment} ($\kappa>1$) \textbf{yields one urn containing almost all balls} and \textbf{sub-preferential attachment} ($\kappa<1$) \textbf{gives rise to rapidly decaying distributions} (no power law);
\item In the case of classical ($\kappa =1$, linear) preferential attachment, we recover power-law tail distribution of urn size \textbf{even when we allow for removing balls from urns picked in proportion to their size, provided the probability of ball extraction remains smaller than or equal to the probability of ball addition} (see \citealp{lansky2014role, cai2011phase} for similar models and similar results). 
\end{itemize}

\begin{rem}
Urn schemes with ball removal have been studied in a variety of models, mostly in the context of growing random networks, see for example 
\cite{thornblad2015asymptotic, moore2006exact, deijfen2010random, cai2011phase, pralat2011edge, cooper2004random, lansky2014role, fenner2005stochastic, janson2004functional}. Other extensions to the classical Yule and Simon models include versions with anti-preferential attachment \citep{deambroggio2020dynamic}, or with node aging \citep{polito2018studies, bertoin2019version, baur2021two} and works trying to make a synthesis of existing models and of their mathematical properties \citep{pachon2016random,chan2003stochastically, krapivsky2001organization, ghoshal2013uncovering}.
\end{rem}

\section{Conclusion}

\cite{Y} had a fantastic intuition when he introduced the pure-birth process, with an approach that was already, as we saw, hinting at the birth-death process and more generally at Markov chains in continuous time. 

We have reviewed some interesting accidents in the history of Yule's heritage: his paternity of the pure-birth process has only been recognized in the late 40's \citep{kendall1949stochastic} after the three most famous papers about birth-death processes \citep{furry1937fluctuation,feller1939grundlagen,kendall1948generalized}  had appeared without citing \cite{Y}; papers in evolutionary biology, even those dealing with \emph{nested} trees (e.g., the multispecies coalescent) cite \cite{Y} for the so-called `Yule tree', simultaneously forgetting that Yule's work is precisely about \emph{nested} birth processes and that it actually never even alludes to the notion of tree; papers in statistical physics and social sciences cite \cite{Y} for introducing a power-law distribution produced by a preferential attachment mechanism, an anachronistic interpretation following its rediscovery by \cite{simon1955class}.

A wealth of other questions of historical interest remain: the relations of G.U. Yule with other statisticians (in particular R.A. Fisher, who like Yule gave a lecture at the 1924 International Congress of Mathematicians in Toronto), his relations with biologists (at the Royal Society and elsewhere) and foremost, the diffusion of his ideas. Indeed, it is difficult to identify  who unearthed \cite{Y} and brought it to the knowledge of D.G. Kendall and his fellows of the Royal Statistical Society in 1948. It seems that even at the time of its publication, \cite{Y} remained unnoticed from mathematicians involved in biological modeling (Laurent Mazliak, pers. comm.): no mention of Yule's paper in Kostitzin's book ``Biologie math\'ematique'' (published in 1937 with a preface by Volterra); no trace of any correspondence between Volterra and Yule in Volterra's archives; and in Mal\'ecot's thesis, the name of Yule is quoted once, on the topic of incomplete dominance in Mendelian genetics... 

In the same issue, \cite{PMacP}  give a wonderful overview of the legacy of \cite{Y} in macroevolutionary studies, showing in particular that its rediscovery by paleontologist D.M. Raup \citep{raup1978cohort,raup1985mathematical} probably came from his reading of \cite{bailey_elements_1964}. Norman T. Bailey was a bio-mathematician educated at Cambridge who had followed lectures by R.A. Fisher and M. Bartlett \citep{armitage2009norman}, from whom he probably had learnt about Yule's work.

As our earlier quotation of Yule's obituary by \cite{kendall1952george} showed, it actually is very unfair that the posterity of Yule virtually uniquely relies on his 1925 paper. However, it is remarkable that this paper, as we have tried to show, remains so modern after a century of existence. Not only has it brought to light the pure-birth process and the computation of its marginal distribution fifteen years ahead of time, it also handles modern notions such as nested branching processes and neutral processes of macroevolution and bears the seeds of contemporary mathematical objects such as random trees, trees within trees and preferential attachment graphs, which intersect, and sometimes reunite, evolutionary biology, statistical physics and probability theory. In view of today's vividness of these questions, it seems that the realm open by \cite{Y} still leaves unanswered deep scientific questions that might well keep us at work for another century... 

Let us finally bring back from this journey a good memory, not only of the ideas of Yule, but of the man himself, by quoting again \cite{kendall1952george}:  
\begin{quote}
``A man of Yule's age has the misfortune of seeing many of his friends and contemporaries precede him to the grave. Some are left to mourn him; far larger is the number of younger men who knew him first as instructor and then as friend and will always remember him as one of the ablest, kindliest and most lovable of men.''
\end{quote}

\vskip6pt

\enlargethispage{20pt}

\paragraph{Acknowledgments.} I am especially grateful to historians of science Laurent Loison (history of biology) and Laurent Mazliak (history of mathematics) for sharing their knowledge of the scientific context in Yule's time. Let me also thank Mike Steel, F\'elix Foutel--Rodier, Corinne Robert and Guillaume Achaz for discussions.  For funding I thank the \emph{Center for Interdisciplinary Research in Biology} (CIRB, Coll\`ege de France) and the \emph{Institute of Biology of \'Ecole Normale Sup\'erieure} (IBENS, \'Ecole Normale Sup\'erieure, Universit\'e PSL).


\vskip2pc

\bibliographystyle{abbrvnat}

%
\bibliography{myrefs} 




\appendix

\section{Proof of Theorem \ref{thm:yule2}}
\label{subsec:joint}

\begin{proof}
We call $\mathcal T$ be the tree of genera and define its \emph{root} as the point of origination of the generation process, at time 0. For simplicity we denote by $N:=N_T^f$ the number of genera sampled at time $T$. See Figure \ref{fig:jccp}(a). 

\begin{figure}[h] \includegraphics[width=\textwidth]{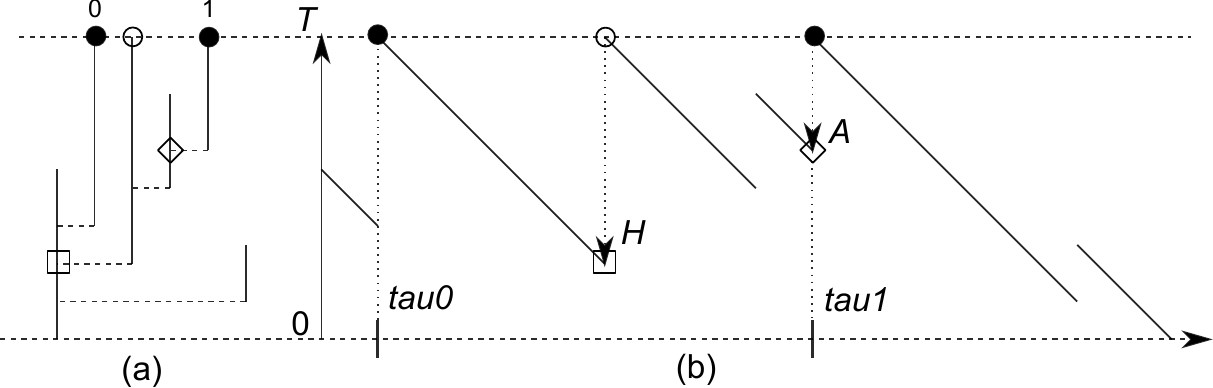} 
 \caption{An oriented tree stopped at time $T$ with $N_T^f=2$ sampled genera and the associated contour process. (a) Out of three extant genera at present time, two are sampled (black dots), one is not (white dot). Coalescence between the two sampled genera is indicated by a square. Birth of genus 1 is indicated by a diamond. (b) Contour process associated with tree of panel (a). Times $\tau_0$ and $\tau_1$ of visits of sampled genus $0$ and $1$ are indicated. Points corresponding to samples, coalescence and birth are reported by the corresponding geometric symbols. At time $T$, genus $1$ has age $A$; coalescence between $0$ and $1$ occurs at depth $H$.} 
 \label{fig:jccp}
 \end{figure}

\paragraph{Contour process.} The so-called \emph{contour process} explores the tree by visiting exactly once each of its points in some order induced by the orientation. 
The exploration starts at the leftmost tip of $\mathcal T$. The value of the contour process at time 0 is thus $\min(z,T)$, where $z$ is the time of death of the primordial genus. The visit of $\mathcal T$ is done by sliding towards the root at speed 1 and the process reports the value $y(t)$ where $y(t)$ is the distance to the root of the point visited at time $t$. Each time the process encounters a birth event, it jumps to the next tip to its right. The process ends when it reaches $0$ (i.e., the root). See Figure \ref{fig:jccp}(b).

\cite{lambert2010contour} showed that under the assumptions [(i)-(ii)-(iii)] on the underlying generation process made earlier, the contour process $X$ has the same law as a \textbf{time-homogeneous, strong Markov process} $\Xi$ killed upon hitting $0$ and reflected below $T$. Indeed, because the generation process is stopped at time $T$, each jump that could make $X$ overshoot $T$ is actually truncated to have $X$ take the value $T$ instead: it will be convenient to consider nevertheless that each jump time is a hitting time of the half-line $(T,\infty)$ as by the process $\Xi$ before reflection. The process $\Xi$ is a piecewise deterministic Markov process, which decreases linearly with slope $-1$ and jumps at rate $g$ with positive jumps distributed as $L$ (in the absence of reflection and killing). We will use repeatedly the fact that $\Xi$ is not only Markovian, it is a L\'evy process \citep{bertoin-Levy} and as such is invariant by translation.

\paragraph{Excursions.} Let $\tau_0, \tau_1, \ldots, \tau_{N-1}$ be the successive times when the process visits a sampled tip. In particular $X_{\tau_i} =T$, but not every time when $X$ hits $T$ is necessarily a $\tau_i$ because of unsampled genera. 
Define $\varepsilon_i$ the $i$-th excursion of $X$ between the visit of two consecutive sampled tips shifted by $-T$, that is:
$$
\varepsilon_i(t) = X_{\tau_{i-1}+t}-T\qquad 0\le t < \tau_i-\tau_{i-1}.
$$
We denote by $V(\varepsilon_i):=\tau_i-\tau_{i-1}$ the lifetime of the $i$-th excursion.
By the strong Markov property the excursions $(\varepsilon_i)$ form a sequence of i.i.d. paths, killed at the first one which hits $-T$. Each of these paths has the same law as 
\begin{itemize}
\item the concatenation of $K$ i.i.d. excursions $(e_j)_{1\le j\le K}$,
\item all distributed as the L\'evy process $\Xi$ starting from $0$ and killed upon hitting $(0,+\infty)$,
\item where $K\ge 1$ is an independent geometric rv with success probability $f$.
\end{itemize}

\paragraph{Ages and coalescence times.} Now set
$$
A_i:= T-X_{\tau_i-}\quad 0=1,\ldots, N-1,
$$
the size of the jump that $X$ makes upon visiting the $i$-th sampled genus, with the convention that $X_{0-}=T$ (to treat the possibility that $\tau_1=0$).
It can be realized that \textbf{$A_i$ is the age of the $i$-th sampled genus} and that the \textbf{coalescence time $H_i$ between genus $i-1$ and genus $i$} can also be retrieved from $X$ as
$$
H_i=T- \inf_{\tau_{i-1}\le t < \tau_{i}} X_t, \quad 1\le i \le N-1.
$$
Now note that
$$
(H_i, A_i)=(-\inf \varepsilon_i, -\varepsilon_i(V(\varepsilon_i)-)),
$$
so that $(H_i,A_i)$ indeed form a sequence of i.i.d. pairs killed at the first value of its first component larger than $T$. Now conditional on $A_i=a$, the number $S_i$ of species in genus $i$ is independent of everything else and it is distributed as the number of species in a genus of age $a$ conditional on the genus to be still extant at $a$, that is, conditional on $Z_a\not=0$, since 0 is absorbing for the process $Z$ and genera can only become extinct if their number of species falls to 0.

Then the i.i.d. property extends to the triples $(H_i,A_i, S_i)$ and it only remains to characterize the common law of the $(H_i, A_i)$.\\
 
Recall the function $W$ defined by \eqref{eqn:integro-diff} and $W_f=1-f+fW$. 
 \cite{lambert2013birth} have proved (Proposition 4.2) that
 $$
W_f(t) = \frac{1}{\PP(-\inf \varepsilon>t)}\qquad t\ge0.
$$
We will also use the following solution to the two-sided exit problem \citep{emery1973exit,takacs1967combinatorial,rogers1990two, bertoin-Levy}
\begin{align}
\label{eqn:scale}
\begin{split}
P_{x}(\Xi \mbox{ hits $(h,+\infty)$ before $0$}) &= 1-\frac{W(h-x)}{W(h)}\qquad 0\le x \le h,
\end{split}
\end{align}
where the index $x$ in $P_x$ indicates that $\Xi_0 =x$.

Informally, it is well-known that the time-reversal of the L\'evy process $\Xi$ is equal in law to $-\Xi$, which can be stated formally as follows. Let $e$ be distributed as the process $\Xi$ started at $0$ and killed at the time $\tau_0^+$ when it hits $(0,+\infty)$. Then conditional on $\{\tau_0^+<\infty, -e(\tau_0^+-)=x\}$, the law of $(-e((\tau_0^+-t)-);0\le t \le \tau_0^+)$ is the same as that of $\Xi$ started at $x$, conditioned and killed upon hitting 0.  

Now recall that each excursion $\varepsilon_i$ can be seen as the concatenation of $K$ i.i.d. excursions $(e_j)_{1\le j\le K}$, so that  
\begin{eqnarray*}
\PP (H<h\,|\, A=x) &=& \PP(-\inf e_j<h, 1\le j \le K\,|\, -e_K(\tau_0^+-)=x)\\
	&=&\sum_{k\ge 1}f(1-f)^{k-1}\PP(-\inf e<h)^{k-1} \PP(-\inf e<h \,|\, -e(\tau_0^+-)=x)\\
	&=&P_x(\tau_0<\tau_h^+\,|\,\tau_0<\infty)f\left(1-(1-f) P_0(\tau_0^+<\tau_h)\right)^{-1}\\
	&=&\frac{W(h-x)}{W(h)}e^{\eta x}f\left(1-(1-f) (1-1/W(h))\right)^{-1}\\
	&=&e^{\eta x}\frac{fW(h-x)}{W_f(h)}
\end{eqnarray*}
where we have used \eqref{eqn:scale} and the fact that $P_x(\tau_0<\infty) = e^{-\eta x}$.
\end{proof}

We now give an alternative proof of the joint law of $(H, A)$ by characterizing the law of $A$ conditional on $H$.
Recall that $v$ is the density of the lifetime $L$ of a genus and set $\bar v(t):= \PP(L>t)$.
\begin{prop}
\label{prop:age}
For any $h>0$,
\begin{equation}
\label{eqn:density-age-2}
P(A\in dx\,|\, H=h)=g \bar v(x) \left(\frac{W_f(h)W'(h-x)}{W'(h)}-fW(h-x)\right)dx\qquad 0\le x<h,
\end{equation}
and
\begin{equation}
\label{eqn:a=h-2}
P(A=h\,|\, H=h) = g\bar v(h) \frac{W_f(h)}{W'(h)}.
\end{equation}
\end{prop}
\begin{rem}
Recalling that $\PP(H>t)=1/W_f(t)$, we can check the agreement between the results of the previous statement and those of Theorem \ref{thm:yule2}. First, differentiating \eqref{eqn:density-age-1} w.r.t. $h$ yields the same result as multiplying  \eqref{eqn:density-age-2} by the density $W_f'/W_f^2$ of $H$, namely (recall $W_f'=fW'$)
\begin{equation*}
P(A\in dx , H\in dh)=\frac{fg \bar v(x)}{W_f(h)} \left(W'(h-x)-\frac{W_f'(h)W(h-x)}{W_f(h)}\right)dx\, dh\qquad 0\le x<h.
\end{equation*}
Second, \eqref{eqn:a=h-1} can also be obtained by multiplying \eqref{eqn:a=h-2} by the density of $H$, namely
\begin{equation*}
P(A=H\in dh) =  \frac{fg\bar v(h)}{W_f(h)} dh.
\end{equation*}
\end{rem}
\begin{proof}
We will use the following fact (see e.g., \citealp{lambert2009allelic}):
\begin{equation}
\label{eqn:random-initial}
P_{L}(\Xi\mbox{ hits $(h,+\infty)$ before $0$}) = \frac{W'(h)}{g W(h)},
\end{equation}
where the index $L$ in $P_L$ indicates that $\Xi_0$ is distributed as the lifetime of a genus, that is, has density $v$.

Let us consider the contour process starting from $T$ (visit of a sampled genus) and stopped after $K$ consecutive visits of T (visit of the next sampled genus in the exploration order), where $K$ is geometric with success probability $f$. 
Conditional on $H=h$, this excursion of the contour process can be decomposed into its parts before and after the unique time $\sigma$ when $X$ makes a jump such that $X_{\sigma-}=T-h$. Since $\sigma$ is a stopping time for $X$ and sampling events are independent and independent of $X$, the law of the post-$\sigma$ part of the excursion (conditional on $H=h$) has the same law as the process $\Xi$ started at the value $U:=(T-h+L)\wedge T$, reflected below $T$ and conditioned on the event $B$ to hit $T$ a geometric number $\tilde K$ of times before hitting $T-h$, where the success probability of $\tilde K$ is $f$. 
If we let $\tau^-$ be the first hitting time of $T-h$ and $\tau^+$ be the first hitting time of $(T,+\infty)$, recalling that $\PP(H>t)=1/W_f(t)$, we get
$$
\PP(B)= \PP_U (\tau^+ <\tau^-) \left(f + (1-f)\left(1-\frac{1}{W_f(h)}\right)\right).
$$
Now thanks to \eqref{eqn:random-initial}, $P_U (\tau^+ <\tau^-)=W'(h)/gW(h)$, so we get
$$
\PP(B)= \frac{W'(h)}{gW(h)}\frac{fW(h)}{W_f(h)}=\frac{fW'(h)}{gW_f(h)}.
$$
Now conditional on $H=h$, $A=H$ if and only if the jump of $X$ at time $\sigma$ is larger than $h$ and $\tilde K=1$, so that
$$
P(A=h\,|\, H=h) = \frac{f\bar v(h)}{\PP (B)} = \frac{g\bar v(h) W_f(h)}{W'(h)},
$$
which proves \eqref{eqn:a=h-2}.
For all other cases, let us distinguish the cases when $\tilde K=1$ and $\tilde K>1$:
\begin{itemize}
\item
Either $\tilde K=1$ and $U<T$. Then the contour process $X$ starts at $U$ and $A= T-X_{\tau^+-}$;
\item Or $\tilde K>1$. Then $A= -\varepsilon_{\tau^+-}$, where $\varepsilon$ is the $\tilde K$-th excursion (corresponding to the next sampled genus), which starts from $0$.
\end{itemize}
Then we have
\begin{align*}
P(A\in dx\,|\, H=h) = \frac{f}{\PP(B)}\int_0^h dz\,v(z)\, P_{T-h+z}(\tau^+ <\tau^-, T-\Xi_{\tau^+-} \in dx) \\
+ \frac{\PP(C)}{\PP(B)} \, P_{T}(\tau^+ <\tau^-, T-\Xi_{\tau^+-} \in dx), 
\end{align*}   
where $C$ is the event that $\tilde K>1$ and all excursions before the last one hit $(T,+\infty)$ before $(T-h)$, so that 
\begin{eqnarray*}
\PP(C) & = & (1-f)\PP_U (\tau^+ <\tau^-) \sum_{k\ge1} f (1-f)^{k-1}\left(1-\frac{1}{W(h)}\right)^{k-1}\\
	&=&(1-f)\frac{W'(h)}{gW(h)}\frac{f}{1-(1-f)\left(1-\frac{1}{W(h)}\right)}\\
	&=&\frac{(1-f)fW'(h)}{gW_f(h)}. 
\end{eqnarray*}
Then
$$
\frac{\PP(C)}{\PP(B)} = 1-f.
$$
Now it is known (see e.g., \citealp{bertoin1997exponential}), that 
$$
P_{T-y}(\tau^+ <\tau^-, T-\Xi_{\tau^+-} \in dx) / dx=g \bar v(x)\left( \frac{W(y)W(h-x)}{W(h)}-\mathbf{1}_{y\ge x} W(y-x)\right).
$$
This yields
\begin{align*}
\frac{P(A\in dx\,|\, H=h)/dx}{g \bar v(x)} = \frac{g W_f(h)}{W'(h)}\int_0^h dz\,v(z)\,  \left(\frac{W(h-z)W(h-x)}{W(h)}-\mathbf{1}_{h-z\ge x} W(h-z-x)\right) \\
+ (1-f)\, \frac{W(0)W(h-x)}{W(h)}. 
\end{align*}   
Defining the convolution product  $M(h) := g \int_0^h dz\,v(z)\, W(h-z)$ and recalling that $W(0)=1$, we get
$$
\frac{P(A\in dx\,|\, H=h)/dx}{g \bar v(x)} = \frac{W_f(h)}{W(h)W'(h)} (M(h)W(h-x) - M(h-x)W(h) ) + (1-f) \frac{W(h-x)}{W(h)}.
$$
Now by \eqref{eqn:integro-diff}, we have $M(h) = g W(h)- W'(h)$,
so that 
\begin{align*}
\frac{P(A\in dx\,|\, H=h)/dx}{g \bar v(x)} 
 = \frac{W_f(h)}{W(h)W'(h)} ((g W(h)- W'(h))W(h-x) - (gW(h-x)- W'(h-x))W(h) ) + (1-f) \frac{W(h-x)}{W(h)}\\
	=\frac{W_f(h)}{W(h)W'(h)}(W'(h-x) W(h) - W'(h) W(h-x))+ (1-f) \frac{W(h-x)}{W(h)}\\
	=\frac{W_f(h)W'(h-x)}{W'(h)} +\frac{W(h-x)}{W(h)}(1-f - W_f(h)),
\end{align*}
that is
$$
P(A\in dx\,|\, H=h)=g \bar v(x) \left(\frac{W_f(h)W'(h-x)}{W'(h)}-fW(h-x)\right)dx\qquad 0\le x<h,
$$
which proves \eqref{eqn:density-age-2}.
\end{proof}

\section{Proofs of Propositions \ref{prop:case1} and \ref{prop:case2}}
\label{app:prop}
\begin{proof}[Proof of Proposition \ref{prop:case1}]
Here, $Z$ is a linear birth-death process with birth rate $\lambda$ and death rate $\mu$. Then it is known \citep{kendall1948generalized} that 
$$
\PP(Z_t=n)=\frac{F'(t)}{\lambda F(t)^2}\left(1-\frac1{F(t)}\right)^{n-1}, 
$$  
where $F(t) =1+\lambda (e^{rt}-1)/r$ when $r:=\lambda-\mu\not=0$ and $F(t) =1+\lambda t$ if $r=0$. The formula can actually be extended in this form to more general $F$, see for example \cite{lambert2010contour} and Remark \ref{rem:geom}. 
Now thanks to Theorem \ref{thm:yule2}, we can integrate this formula against the law of $A$ w.r.t. $t$ to get the law of the number $S$ of species per genus:
\begin{align}
\label{eqn:p(s=n)}
\begin{split}
\PP(S=n)&=\int_0^\infty g e^{-\eta t}\frac{F'(t)}{\lambda F(t)^2}\left(1-\frac1{F(t)}\right)^{n-1}dt.
\end{split}
\end{align}
First assume $\lambda< \mu$, so that $\lim_{t\to\infty}F(t) = -\mu/r$. If $\eta=0$ (i.e., $g=g^\star$), we can integrate \eqref{eqn:p(s=n)} to get
$$
\PP(S=n)=\frac{g}{\lambda}\frac{(\lambda/\mu)^n}{n}.
$$ 
If $\eta >0$ (i.e., $g>g^\star$), an integration by parts yields
\begin{equation*}
\PP(S=n)= \frac g{\lambda n}\int_0^\infty \eta  e^{-\eta t}\left(1-\frac1{F(t)}\right)^{n}dt.
\end{equation*}  
Now $F$ is strictly increasing so that $F(t) <F(\infty)= \mu/r$. As a consequence,
$$
\PP(S=n)< \frac{g}{\lambda n} \int_0^\infty \eta e^{-\eta t}(\lambda/\mu)^n dt = \frac{g}{\lambda n}(\lambda/\mu)^n.
$$ 
If $\lambda\ge \mu$, then $1/F$ is a decreasing bijection from $[0,\infty)$ to $(0,1]$. Let $G:(0,1]\to [0,\infty)$ be its inverse. Then we can change variable in \eqref{eqn:p(s=n)}, to get
$$
\PP(S=n)=\frac g\lambda \int_0^1  e^{-\eta G(x)}\left(1-x\right)^{n-1}dx.
$$
To obtain an expression for $G$, we write $x=1/F(G(x))$. 
Let us first assume that $\lambda=\mu$, so that $x=(1+\lambda G(x))^{-1}$ and
$$
G(x) = \frac{1-x}{\lambda x}.
$$
Plugging this into the integral,  
$$
\PP(S=n)=\frac g\lambda \int_0^1 e^{-\eta(1-x)/(\lambda x)} \left(1-x\right)^{n-1}dx =\frac {g e^{\eta/\lambda}}\lambda \int_0^1 R_{n-1}(x) dx,
$$
where $R_{n-1}(x)=e^{-c/x} \left(1-x\right)^{n-1}$, with $c= \eta/\lambda$. Now $R_n'(x) = e^{-c/x} \left(1-x\right)^{n-1}(c(1-x) - nx^2)/x^2$. The function $R_n'$ has only one positive root equal to 
$x_n= (-c+\sqrt{c^2+4nc})/2n\sim \sqrt{c/n}$ when $n$ is large. Then $R_n$ achieves its maximum at $x_n$ and
$$
\ln R_n(x_n) = -\frac c{x_n}+n\ln (1-x_n) \sim -2\sqrt{cn}.
$$
By Laplace's method, we get $\ln(\PP(S=n))/\sqrt n$ converges to $-2\sqrt c$, which yields \eqref{eqn:lap}.
Now assume $\lambda>\mu$, so that $x=(1+\lambda (e^{rG(x)}-1)/r)^{-1}$. Then we get
$$
e^{rG(x)} = 1+\frac r\lambda\frac{1-x}{x} = \frac{r+\mu x}{\lambda x}\qquad x\in (0,1],
$$
and
$$
\PP(S=n)=\frac g\lambda \int_0^1  \left(\frac{\lambda x}{r+\mu x}\right)^{\eta/r} \left(1-x\right)^{n-1}dx.
$$
Thus we get the double inequality
$$
\frac g\lambda \int_0^1  x^{\eta/r} \left(1-x\right)^{n-1}dx\ \le\  \PP(S=n)\ \le\ \frac g\lambda \int_0^1  \left(\frac{\lambda x}{r}\right)^{\eta/r} \left(1-x\right)^{n-1}dx.
$$
Setting $\alpha = \eta/r$, we can conclude that $(\PP(S=n)n^{\alpha+1})$ is bounded and bounded away from 0, which yields \eqref{eqn:log-log}.
 \end{proof}

\begin{proof}[Proof of Proposition \ref{prop:case2}]
Let us first prove Equations \eqref{eqn:case2-eq1} and \eqref{eqn:case2-eq2}. For $k\ge 1$, let $J_k$ be the first hitting time of $k$ by the general pure-birth process $Z$:
$$
J_k:=\inf\{t>0: Z_t=k\},
$$
so that $J_1=0$ and
$$
\PP (S=k, A\in da)= \PP (J_k<a< J_{k+1})\,ge^{-ga} da.
$$

Classical arguments using competing exponential clocks show that the sequence of random variables $(A\wedge J_{k+1} - J_k)_{k\ge 1}$ killed at its first negative value (excluded) has the same law as the sequence $(C_{k}\wedge D_k)_{k\ge 1}$ killed at the first $k$ (included) such that $D_k>C_k$, where the $(D_k)$ are i.i.d. exponential random variables with parameter $g$ and the $(C_k)$ are independent exponential random variables with parameter $\lambda_k$.
In addition, another classical argument shows that $C_k\wedge D_k$ is an exponential variable with parameter $g+\lambda_k$ independent of the event $\{D_k<C_k\}$, which has probability $g/(g+\lambda_k)$.
As a consequence, we first get 
$$
\PP (S=n)= \PP(C_n<D_n)\prod_{k=1}^{n-1} \PP(C_k<D_k)= \frac{g}{g+\lambda_n}\prod_{k=1}^{n-1} \frac{\lambda_k}{g+\lambda_k},
$$
which is \eqref{eqn:case2-eq1}.
Second, conditional on $S=n$, $A=A-J_n +\sum_{k=1}^{n-1}(J_{k+1}-J_k)$ has the same law as $C_n\wedge D_n + \sum_{k=1}^{n-1} C_k\wedge D_k$ which yields \eqref{eqn:case2-eq2}.

We now prove the second part of Proposition \ref{prop:case2} by repeatedly using \eqref{eqn:case2-eq1}.
First assume that $(\lambda_n)$ is bounded. Then there is $c>0$ such that $g+\lambda_n\le c$, and because $\ln(1-x)\le -x$,
\begin{eqnarray*}
\ln \big(\PP (S=n)\big)&=& \ln \left(\frac{g}{g+\lambda_n}\right)+\sum_{k=1}^{n-1} \ln \left(1-\frac{g}{g+\lambda_k}\right)\\
	&\le & -\sum_{k=1}^{n-1}\frac{g}{g+\lambda_k}\le -(n-1)g/c,
\end{eqnarray*}
which proves that the sequence $(\PP (S=n))$ is exponentially decaying.

Next assume there are $0<\delta\le \kappa <1$, some positive constants $c_0, c_1$ and some integer $n_0$ such that $c_0n^{\kappa}\le g+\lambda_n\le c_1n^{\delta}$ and $\lambda_n\ge g$ for all $n\ge n_0$. 
Using the double inequality $-2x\le \ln(1-x)\le -x$ which holds for any $x\in [0,1/2]$,
\begin{equation*}
C_0-\sum_{k=n_0}^{n-1}2gc_0k^{-\kappa} \le \ln \big(\PP (S=n)\big) -\ln \left(\frac{g}{g+\lambda_n}\right)\le C_1-\sum_{k=n_0}^{n-1}gc_1k^{-\delta}.
\end{equation*}
As a consequence, 
$$
C_0(g+\lambda_n)e^{-c_0n^{1-\kappa}}\le \PP (S=n)\le C_1(g+\lambda_n)e^{-c_1n^{1-\delta}},
$$
where the $c$'s are (other) positive constants, so that $(\PP (S=n))$ has a heavy tail but no power law tail.

Next assume that $\lambda_n\ge cn^{\delta}$ for some $\delta>1$ and $c>0$. Then the number of species in any genus $i$ is a nondecreasing process $(Z^{(i)})$, which equals $n$ after a time with expectation smaller than $\sum_{k=1}^n 1/(ck^{\delta})$ which converges to some finite value as $n\to\infty$. This shows that in genus $i$ the number of species blows up at some finite time $\sigma_i$, meaning that $\lim_{t\uparrow \sigma_i }Z_t^{(i)}=+\infty$. The minimum of all $\sigma_i$'s is realized for some unique $\sigma_J$, so that the total number of species summed over all genera remains finite for all $t<\sigma_J$ while the number of species in genus $J$, and in genus $J$ only, blows up at time $\sigma_J$. It results in the fraction of all species that are present in genus $J$ going to 1 as $t\uparrow \sigma_J$ (or equivalently, as the total number of species goes to infinity).

Finally assume that $\lim_n\lambda_n/n=s\in(0,\infty)$. Now using the fact that $-\ln(1-x) -x\le x^2/2$ for any $x\in [0,1/2]$, there are constants $C$ and $C'$ such that
$$
0\le 	-\ln \big(\PP (S=n)\big)+\ln \left(\frac{g}{g+\lambda_n}\right)-\sum_{k=1}^{n-1} \frac{g}{g+\lambda_k}\le \frac12\sum_{k=1}^{n-1} \left(\frac{g}{g+\lambda_k}\right)^2+C\le C'
$$
Now
$$
\left| \ln \left(\frac{g}{g+\lambda_n}\right)-\sum_{k=1}^{n-1} \frac{g}{g+\lambda_k}-\ln \left(\frac{g}{sn}\right)+\sum_{k=1}^{n-1} \frac{g}{sk}\right|\le \left| \ln \left(\frac{sn}{g+\lambda_n}\right)\right| +\frac gs\sum_{k=1}^{n-1}\frac{\left|1-(sk)/(g+\lambda_k)\right|}{k}.
$$
Now let $\varepsilon >0$. Because $\lambda_n/(sn)\to 1$, there is some (other) constant $C$ such that for all $n$,
$$
\left| \ln \left(\frac{g}{g+\lambda_n}\right)-\sum_{k=1}^{n-1} \frac{g}{g+\lambda_k}+\ln n+\sum_{k=1}^{n-1} \frac{g}{sk}\right|\le C+\varepsilon \ln n.
$$
Putting everything together shows that 
$$
\left| \ln \big(\PP (S=n)\big)+ \ln n +\frac gs \ln n \right|\le C'+\varepsilon\ln n
$$
for some (other) constant $C'$. Finally, this can be expressed as
$$
\lim_{n\to\infty}\frac{\ln \big(\PP (S=n)\big)}{\ln n} = -1 -\frac gs,
$$
which ends the proof of Proposition \ref{prop:case2}.
\end{proof}

\end{document}